\documentclass[journal]{IEEEtran}
\usepackage{amsmath,amsfonts,amssymb}
\usepackage{array}
\usepackage{booktabs}
\usepackage{multirow}
\usepackage{graphicx}
\graphicspath{{Figures/}}
\usepackage[caption=false,font=normalsize,labelfont=sf,textfont=sf]{subfig}
\usepackage{textcomp}
\usepackage{stfloats}
\usepackage{url}
\usepackage{cite}

\usepackage{siunitx}
\DeclareSIUnit{\year}{yr}
\usepackage{physics}
\usepackage{xspace}
\usepackage[hidelinks]{hyperref}
\usepackage[switch]{lineno}

\newcommand{\dshift}{\textsc{D-SHIFT}\xspace}
\DeclareRobustCommand{\rotplus}{\texorpdfstring{\raisebox{0.15ex}{\rotatebox[origin=c]{45}{+}}}{+}}

\begin{document}

\title{D-SHIFT: Transferring High Spatial Information from GRACE Monthly TWSA Mascon to Daily Products Using Generative Adversarial Networks}

\author{Andreas~Dombos$^1$,
        Junyang~Gou$^{1,2,*}$,
        and~Benedikt~Soja$^1$%
\thanks{$^{1}$Institute of Geodesy and Photogrammetry, ETH Zurich, 8093 Zurich, Switzerland.}%
\thanks{$^{2}$Department of Earth, Atmospheric and Planetary Sciences, Massachusetts Institute of Technology, Cambridge, MA 02139, USA.}%
\thanks{Corresponding author: Junyang Gou (e-mail: jungou@ethz.ch).}}


\maketitle

\begin{abstract}
The Gravity Recovery and Climate Experiment (GRACE) and GRACE Follow-On missions provide monthly terrestrial water storage anomaly (TWSA) estimates for monitoring large-scale water storage change. The monthly temporal resolution of official products limits the analysis of high-frequency hydrological events, while existing daily GRACE products often have reduced spatial resolution due to sparse groundtrack coverage and required smoothing and regularization. This study introduces \dshift{} (Daily Spatial High-Resolution Inference via Feature Transformation), a deep learning-based framework for generating daily, high-resolution TWSA fields from daily spherical harmonic coefficient (SHC) solutions. The model is trained in the monthly domain by using low-resolution daily solutions and other auxiliary features as inputs, while targeting on monthly mascon products. The model is then applied to daily SHC inputs to generate products with similar spatial resolution of monthly products. Monthly validation against mascon products gives a global mean root mean square error of about \SI{2.3}{\centi\meter}, with good correlation and explained variance agreement. Daily analyses show that \dshift{} produces spatially coherent day-to-day fields and improves basin-scale trend and seasonality estimates compared with low-resolution SHC. The basin-area double-difference analysis indicates that these gains are most relevant for spatially localized signals affected by smoothing and leakage. In Greenland, \dshift{} better reproduces coastal mass-loss patterns and gives a basin-mean trend of \SI{-10.5}{\centi\meter\per\year}, close to the CSR Monthly value of \SI{-12.0}{\centi\meter\per\year}.
\end{abstract}



\section{Introduction}
\IEEEPARstart{T}{he} Gravity Recovery and Climate Experiment (GRACE) and its successor, GRACE Follow-On (GRACE-FO), have provided an unprecedented record of the Earth's monthly mass redistribution from 2002 to 2017, and since 2018, respectively~\cite{tapley2004GracePrinciple,wahr2004GracePrinciple,landerer2020GRACE-FO}. These twin-satellite missions measure Earth's gravity field variations, enabling the estimation of changes in terrestrial water storage anomalies (TWSA), representing the integrated vertical column of water stored in all forms, including soil moisture, snow, ice, surface water bodies, groundwater, etc.~\cite{Chen2022}. GRACE-derived products have reshaped large-scale land-hydrology assessments, supporting basin-scale analyses of water availability, long-term drying/wetting trends, and the hydrological imprint of climate variability~\cite{rodell2018EmergingTrend,rodell2023water}. However, GRACE products are inherently limited by their coarse spatial resolution and monthly temporal resolution \cite{hess-22-2867-2018,humphrey2023Review,Gou2024}. These constraints arise from the spatio-temporal sampling limits of low-Earth orbit satellites and the necessary post-processing steps \cite{KURTENBACH201239, hess-22-2867-2018, Landerer2012, Wahr2006}. While these products have proven adequate for studying large-scale, long-term hydrological trends, the constraints pose significant challenges for applications requiring finer temporal and spatial details~\cite{vishwakarma2018GRACEResolution,Gou2024}. As a result, sub-monthly variations are hardly captured in monthly fields, limiting their usability for flash flood/drought monitoring, early warning, and regional water management, which require finer spatial detail and higher sampling frequency \cite{Chen2022}.

To address the aforementioned difficulties, some efforts have been made to generate high-temporal-resolution GRACE products, including products derived from daily spherical harmonic coefficients (SHC)~\cite{https://doi.org/10.1029/2019JB017415, https://doi.org/10.5880/icgem.2018.003}, related Kalman filter/smoother variants \cite{KURTENBACH201239, hess-22-2867-2018} and daily or 5-day mascon solutions designed to expose sub-monthly variability \cite{Croteau2020, Rateb2024}. However, these products face fundamental challenges: (i) rely on stochastic/dynamic constraints (e.g., Kalman filtering, autoregressive stabilization) that couple consecutive estimates, so the effective temporal resolution is lower than the nominal daily resolution; and (ii) leakage/smoothing effects that can attenuate amplitudes in smaller or rapidly varying basins. Consequently, existing high-temporal-resolution solutions often have lower spatial resolution which results in misrepresenting basin-scale extremes or phase in dynamic regions \cite{hess-22-2867-2018, KURTENBACH201239}. Machine-learning methods have already been used at monthly scale to combine GRACE observations with ancillary data and recover finer spatial structure \cite{Sun2019, essd-11-1153-2019, Li2019, Gou2024, gou2025DSOBP}. These monthly-scale results motivate data-driven complements to classical estimation also for daily products. Among deep-learning approaches, generative adversarial networks (GANs; \cite{goodfellow2020GAN}) are attractive because they can map coarse inputs to sharper outputs in super-resolution settings \cite{goodfellow2014generative, rakotonirina2020esrgan+, Wang2023}, thus, opening new avenues for refining GRACE daily data.

In this study, we introduce \dshift{} (Daily Spatial High-Resolution Inference via Feature Transformation), a monthly-to-daily transfer framework based on a modified nESRGAN+ architecture~\cite{rakotonirina2020esrgan+}. The model is trained only in the monthly domain, where paired low-resolution SHC-based fields and high-resolution mascon fields are available, and is then applied to daily SHC-based products without additional retraining or fine-tuning. Auxiliary predictors from GLDAS and ERA5 provide meteorological and land-surface context during both monthly training and daily application~\cite{https://doi.org/10.5067/txbmlx370xx8, https://doi.org/10.24381/cds.adbb2d47}. Daily estimates produced in this way inherit spatial characteristics of the monthly reference while retaining high-frequency variability from the daily inputs. They should therefore be interpreted as model-based daily estimates rather than as temporal interpolation. The remainder of the manuscript is structured as follows. Section~\ref{sec:datamethods} describes the data, preprocessing, model architecture, and training setup. Section~\ref{sec:results} presents the monthly and daily evaluations, and Section~\ref{sec:conclusions} summarizes the main findings and limitations.

\section{Data and Methods}\label{sec:datamethods}

\begin{figure}[!htb]
    \centering
    \includegraphics[width=\columnwidth]{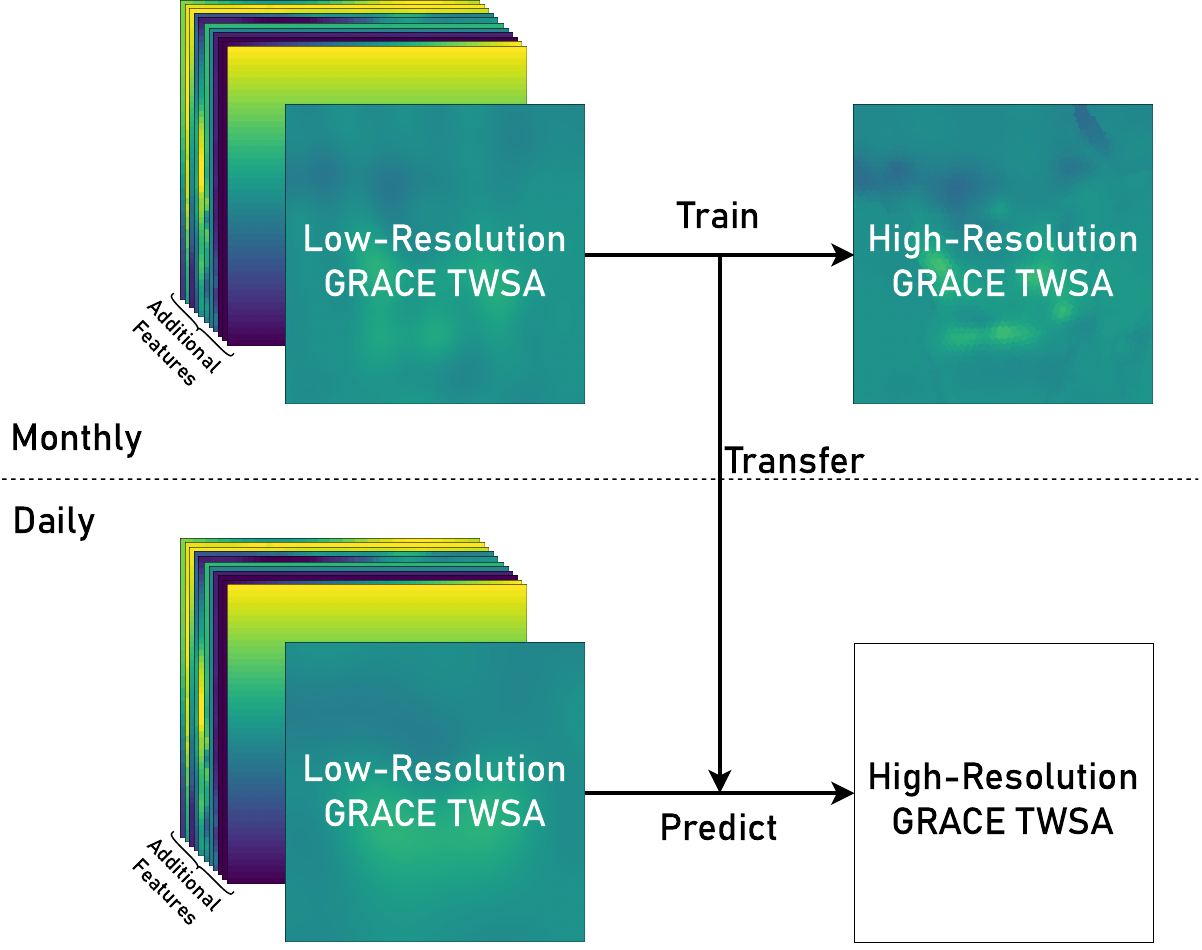}
    \caption{Block diagram of the two-stage \dshift{} pipeline: (i) monthly training, in which the generator is optimized to map low-resolution monthly GRACE inputs and auxiliary predictors (GLDAS, ERA5) to high-resolution monthly CSR mascon targets; and (ii) daily application, in which the trained generator parameters are reused to infer high-resolution daily TWSA from daily inputs. ``Additional Features'' denote the auxiliary meteorological and hydrological variables. The square tiles in the schematic represent the patch-wise processing used at \SI{0.25}{\degree} grid spacing.}
    \label{fig:method}
\end{figure}

Figure~\ref{fig:method} summarizes the workflow of this study. During training, low-resolution monthly SHC-based inputs together with auxiliary predictors are mapped to high-resolution monthly mascon targets. The trained model is then applied to daily SHC-based inputs and daily auxiliary predictors to infer daily high-resolution TWSA. No additional learning step is required in the daily domain.

\subsection{GRACE(-FO) data}\label{sec:gracedata}
\subsubsection{Monthly SHC-based solution from ITSG}
The ITSG-Grace2018 \cite{https://doi.org/10.1029/2019JB017415, https://doi.org/10.5880/icgem.2018.003} release provides unconstrained monthly SHC products up to degree/order of 120. The processing steps following RL06 definitions. The resulting monthly fields represent hydrosphere and cryosphere signals associated with solid-Earth effects (incl. GIA and earthquakes), and any residual atmosphere--ocean variability not removed by the dealiasing/tidal models \cite{https://doi.org/10.1029/2019JB017415}. To align the spatial resolution of monthly products to their daily solutions, we truncated the SHC products to ddgree/order 40 and smoothed with a \SI{300}{\kilo\meter} Gaussian filter so that the monthly inputs have a spatial scale comparable to the daily ITSG inputs used by the generator during daily inference. The filtered fields are generated on \SI{0.25}{\degree} grid to match the CSR mascon products.

\subsubsection{Monthly mascon solution from CSR}
The CSR RL06.2 Mascon solutions provide estimates of TWSA at one-degree hexagon mascons with a nominal grid spacing of $\SI{0.25}{\degree} \times \SI{0.25}{\degree}$ to better represent coastlines~\cite{save2020csr}. Estimation uses Tikhonov regularization with parameters derived purely from GRACE/GRACE-FO data without external constraints and does not require empirical destriping/smoothing processing \cite{save2020csr, Save2016}. In this way, the mascon solutions can have a bit higher effective spatial resolution than the SHC-based solutions~\cite{scanlon2016global}. The other classical corrections, including GIA corrections, stay the same with RL06 convention. Therefore, the final Level-3 (L3) products are directly comparable to the SHC-based solutions. These solutions serve as the high-resolution monthly ground truth for this study.

\subsubsection{Daily SHC-based solution from ITSG}
The ITSG-Grace2018 model also includes daily gravity field solutions, available as SHC up to degree and order 40 \cite{https://doi.org/10.5880/icgem.2018.003, https://doi.org/10.1029/2019JB017415}. These have been stabilized using an autoregressive model, derived by fitting coefficients to the dealiasing error estimates contained in the ESA ESM and detrended LSDM residuals \cite{https://doi.org/10.5880/icgem.2018.003}. Although the ITSG project also provides pre-gridded daily products, we generated own gridded dataset from the filtered SHC to ensure uniform processing steps across daily and monthly domains. To maintain consistency with our low-resolution monthly dataset, the daily spherical harmonics are smoothed with the same \SI{300}{\kilo\meter} Gaussian filter. The filtered data are then resampled from $\SI{0.5}{\degree} \times \SI{0.5}{\degree}$ to $\SI{0.25}{\degree} \times \SI{0.25}{\degree}$, matching the CSR Mascon grid. The daily solutions serve purely as input features during inference, and are never used during the training process.

\subsection{Auxiliary data}\label{sec:auxdata}
\subsubsection{GLDAS}

The Global Land Data Assimilation System (GLDAS; \cite{https://doi.org/10.5067/txbmlx370xx8}) combines satellite observations and in-situ measurements to estimate land surface states and fluxes. This study uses TWS outputs from GLDAS-2.2 driven by the Catchment Land Surface Model (CLSM) and includes data assimilated from GRACE, with baseline values for 2004.000--2009.999 removed to align TWSA definitions. GLDAS data are available in daily and monthly forms at $\SI{0.25}{\degree} \times \SI{0.25}{\degree}$ resolution \cite{https://doi.org/10.5067/txbmlx370xx8,rui22}.

\subsubsection{ERA5}

ERA5, the fifth-generation ECMWF reanalysis, provides climate and meteorological data on an hourly basis, aggregated here to daily and monthly scales at $\SI{0.25}{\degree} \times \SI{0.25}{\degree}$ resolution \cite{https://doi.org/10.24381/cds.adbb2d47}. Key parameters include temperature, wind components, as well as precipitation ($P$), evapotranspiration ($ET$), and runoff ($R$). These variables inform the water balance equation \cite{Chen2022, lehmann2022well}:
\begin{equation}
    \dv{\mathrm{TWS}}{t} = P - ET - R,
\end{equation}
linking hydrological and meteorological parameters to changes in terrestrial water storage. The ERA5 variables are included to enhance the model's capacity to account for regionally varying hydrological responses. Similar to the TWSA standard, baseline values from 2004.000--2009.999 were individually removed for each variable.

\subsubsection{HydroSHEDS Basins}\label{sec:basins}

HydroSHEDS supplies detailed global river basin delineations, with the Level 3 dataset defining 292 sub-basins \cite{Lehner2013}. These basin boundaries fulfill two main purposes. First, they support the CutMix augmentation strategy by orienting data augmentation toward hydrologically significant features (see Section~\ref{sec:model}). Second, they enable basin-scale evaluations of the model's ability to replicate TWSA in diverse hydrological regimes (see Section~\ref{sec:eval}).

\subsection{Feature Preprocessing and Training Strategy}\label{sec:preproc}
Preprocessing steps are needed to standardize input features with various physical units to ensure the consistency of input magnitudes, thereby optimizing its suitability for model training~\cite{Goodfellow-et-al-2016}. This process includes data normalization, patch extraction, a pixel-based prediction framework, and the division of data into training and validation subsets. To normalize each dataset to the range $[0, 1]$, the following min-max scaling \cite{Goodfellow-et-al-2016, Garca2014} is applied:
\begin{equation}
x_{\text{norm}} = \frac{x - x_{\text{min}}}{x_{\text{max}} - x_{\text{min}}},
\end{equation}
where $x$ is the original data value, $x_{\text{min}}$ is the minimum value of the dataset, and $x_{\text{max}}$ is the maximum value of the dataset. This procedure guards against any single feature disproportionately influencing the training process \cite{Goodfellow-et-al-2016, Garca2014}. A summary of the minimum and maximum values for each feature is provided in Table~\ref{tab:feature_limits}.

After normalization, the global features are divided into patches with a size of $\SI{12}{\degree} \times \SI{12}{\degree}$, equivalent to $48 \times 48$ pixels at $0.25^\circ$ sampling resolution. This patch size is chosen to accommodate multiple objectives in a single configuration: First, it is large enough to capture GRACE's monthly effective spatial resolution \cite{hess-22-2867-2018, Croteau2020}, which spans approximately \SIrange{300}{400}{\kilo\meter} (roughly \SIrange{3}{4}{\degree} at the equator), thereby encompassing hydrological features such as river basins and large-scale water storage patterns. Second, extending well beyond GRACE resolution reduces edge effects by preventing boundary pixels from dominating the training and prediction processes. The increased spatial coverage helps the model learn broader hydrological interactions that smaller patches might miss. Although the physical distance of one degree of longitude diminishes at higher latitudes, $\SI{12}{\degree} \times \SI{12}{\degree}$ remains sufficient to capture key hydrological signals relevant over land. This patch size also balances computational efficiency with the need to retain important regional characteristics. Global predictions are generated one pixel at a time. For each target pixel, a $48 \times 48$ patch is extracted so that the target pixel is centered within this patch. The model's prediction corresponds directly to the central pixel in each extracted patch, ensuring that local context informs each estimate. By applying this approach across the entire domain, the final global map is compiled without introducing discontinuities or border artifacts. The monthly dataset (all months from 2003 to 2017 with overlapping data availability) is split into a training subset and a validation subset. Specifically, the years 2008, 2015, and 2016 are excluded from training and treated as a monthly validation set to (i) tune hyperparameters and (ii) verify that the learned mapping generalizes to periods with different noise characteristics at the monthly scale, while all remaining years form the monthly training set. We intentionally avoid calling this held-out subset a test set because it is used for hyperparameter tuning and is therefore not fully unseen in a strict evaluation sense. Rather, the main objective of \dshift{} is to assess robustness at the monthly scale before applying the model to the daily domain, where no direct ground truth exists. After model selection, the final model is retrained on all years (including 2008, 2015, and 2016) before being applied to daily inputs.

\begin{table}[!t]
\caption{Minimum and Maximum Values Used for Feature-Wise Min--Max Scaling to [0,1] Before Training/Inference (Daily/Monthly Aggregates as Described in Sections~\ref{sec:gracedata} and~\ref{sec:auxdata}). Units are Shown per Variable.\label{tab:feature_limits}}
\centering
\begin{tabular}{lcc}
\toprule
\textbf{Feature}       & \textbf{Min}                 & \textbf{Max}                \\
\midrule
GRACE Output [\si{\centi\meter}]  & $-1.126 \times 10^3$         & $4.923 \times 10^2$         \\
GRACE Input [\si{\centi\meter}]   & $-1.876 \times 10^2$         & $7.518 \times 10^1$         \\
GLDAS [\si{\centi\meter}]         & $-2.332 \times 10^2$         & $2.076 \times 10^2$         \\
Temperature [\si{\kelvin}]    & $-4.036 \times 10^1$         & $3.346 \times 10^1$         \\
u-Wind [\si{\meter\per\second}]       & $-1.273 \times 10^1$         & $1.280 \times 10^1$         \\
v-Wind [\si{\meter\per\second}]      & $-1.211 \times 10^1$         & $1.100 \times 10^1$         \\
ET [\si{\meter\per\second}]           & $-5.029 \times 10^{-4}$      & $3.161 \times 10^{-4}$      \\
R [\si{\meter}]              & $-2.344$                          & $3.674$         \\
P [\si{\meter}]             & $2.310 \times 10^{-3}$                          & $3.590 \times 10^{-3}$      \\
\bottomrule
\end{tabular}
\end{table}

\subsection{Model Architecture}\label{sec:model}

\begin{figure*}[!t]
    \centering
    \includegraphics[width=0.8\textwidth]{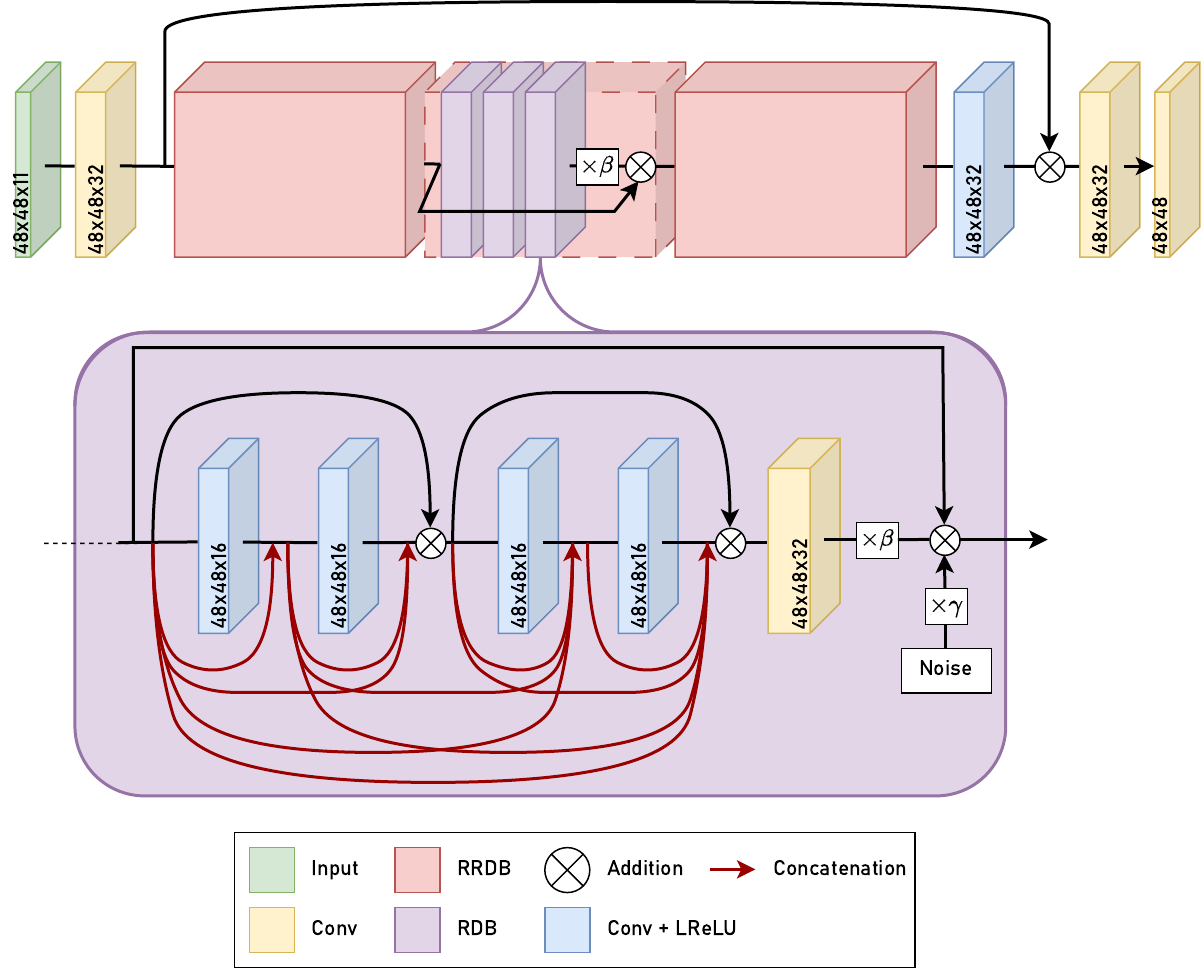}
    \caption{Generator architecture (modified nESRGAN+) used in \dshift{}. A $48\times48$ context patch (at $0.25^\circ$ sampling) is processed by an initial convolution, followed by a stack of RRDBs (Residual-in-Residual Dense Blocks). Inside each RRDB, dense blocks (RDBs) are connected by concatenations (red arrows) and residual additions (\rotplus); the residual branch is scaled by a factor $\beta$ before being added to the skip connection (residual scaling). Gaussian noise is injected after RRDBs and scaled by $\gamma$ (noise amplitude scaling). Final convolutions map the feature tensor back to a single-channel TWSA output. Example tensor shapes are annotated in the diagram.}
    \label{fig:generator}
\end{figure*}

This study employs a modified version of the Enhanced Super-Resolution Generative Adversarial Network Plus (nESRGAN+; \cite{rakotonirina2020esrgan+, https://doi.org/10.48550/arxiv.1809.00219}) that is specifically adapted to address the challenges of enhancing the temporal resolution of GRACE-derived TWSA datasets. Although nESRGAN+ is traditionally recognized for its skills in super-resolution tasks \cite{rakotonirina2020esrgan+}, the adapted model (see Figure~\ref{fig:generator}) builds upon nESRGAN+ with minor modifications that are essential for processing global GRACE TWSA. Residual-in-Residual Dense Blocks (RRDBs) serve as the foundational component of the generator network, integrating dense connections in residual learning structures \cite{https://doi.org/10.48550/arxiv.1809.00219}. This design promotes robust feature extraction by enabling the model to capture both small-scale spatial details and large-scale mass variations. The inclusion of dense connections further alleviates the vanishing gradient problem commonly encountered in deep architectures~\cite{huang2017densely}, thereby ensuring stable training over extensive networks. Another important component is the injection of Gaussian noise after each residual dense block, which introduces controlled stochastic variability that simulates natural fluctuations in hydrological systems \cite{rakotonirina2020esrgan+, Goodfellow-et-al-2016}. This controlled noise avoids over-smoothing and preserves important localized anomalies.

\begin{figure*}[!t]
    \centering
    \includegraphics[width=0.8\textwidth]{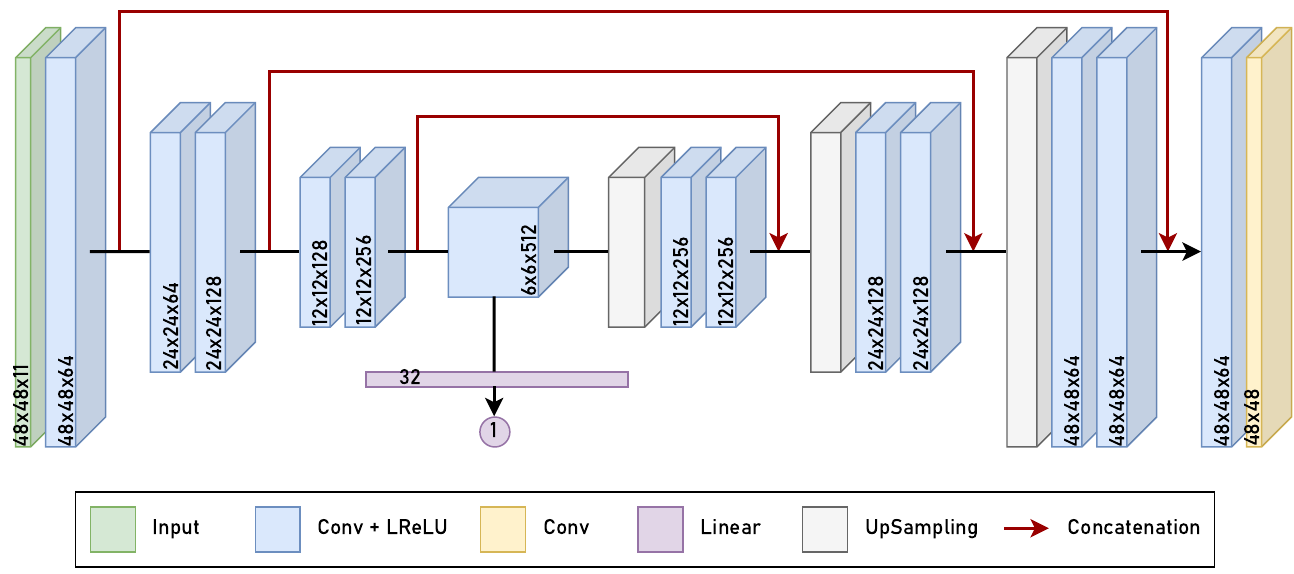}
    \caption{The U-Net--style discriminator used in \dshift{} producing two authenticity assessments: a global score from the encoder pathway (the purple output) and a pixel-wise map from the decoder pathway (the yellow output). During training, basin-aware CutMix composites (Section~\ref{sec:model}) focus the discriminator on hydrologically meaningful regions, while a consistency regularizer enforces stable responses under mask mixing. This design improves sensitivity to both basin-scale structures and fine spatial detail.}
    \label{fig:discriminator}
\end{figure*}

Unlike standard GAN discriminators that primarily evaluate images as a whole, the discriminator in this model has been modified to address the specific requirements of TWSA data: It integrates a U-Net-based architecture \cite{schonfeld2020u}, which is capable of performing both pixel-wise and global evaluations, thereby enabling high-resolution assessments of similarities between the generated patches and real hydrological signals (see Figure~\ref{fig:discriminator}). The introduction of the CutMix augmentation~\cite{schonfeld2020u, https://doi.org/10.48550/arxiv.1905.04899} further advances this functionality by combining real and generated images according to basin-specific masks obtained from the HydroSHEDS dataset. This basin-based augmentation forces the discriminator to focus on unique hydrological features, allowing it to more effectively identify discrepancies at both local and basin-wide scales.

Although nESRGAN+ is a powerful architecture, processing large-scale global datasets demands careful attention to computational efficiency. Our adaptation reduces the total number of RRDBs, thereby reducing the number of trainable parameters and achieving faster training and inference times, while still leaving enough model capacity to represent multiscale hydrological processes. Rather than operating on entire global data, the model processes smaller, overlapping patches. This strategy diminishes memory requirements and retains local consistency, while a subsequent post-processing stage consolidates these patches into coherent global outputs, as described in Section~\ref{sec:preproc}. Basin-aware CutMix in the discriminator focuses part of the adversarial training on hydrologically meaningful spatial units, so discrepancies across basin boundaries are penalized more directly than with random mask mixing. It also benefits from the auxiliary datasets described in Section~\ref{sec:auxdata}.

\subsection{Model Training and Monthly-to-Daily Transfer in \dshift{}}

Refining GRACE-derived TWSA data from monthly to daily scale presents a significant challenge because globally available daily high-resolution ground truth does not exist. The model developed in this study is trained solely on monthly data and then applied to daily inputs, transferring the spatial relationship learned between monthly inputs and targets to daily scales. This strategy therefore relies on monthly-to-daily transfer rather than additional optimizing in the daily domain: relationship learned from the monthly domain is reused during inference to estimate daily variations.

The training processes start with a pretraining phase aimed at initializing the generator \cite{https://doi.org/10.48550/arxiv.1809.00219}. During this phase, the generator minimizes an $L_1$-based loss function. Compared with an $L_2$ loss, this choice is less sensitive to a small number of large residuals and typically preserves sharper spatial transitions:
\begin{equation}
    L_{G,1} = \mathbb{E}_{x}\bigl[\| x_f - x_r \|_1\bigr],
\end{equation}
where $\mathbb{E}_{x}$ is the expectation over a batch, $\|\cdot\|_1$ denotes the $L_1$-norm, $x_f$ is the generated output, and $x_r$ denotes the ground truth. The goal of this pretraining is to ensure pixel-level alignments between synthesized and real data, thereby reducing the complexity of the subsequent adversarial training stage. This approach helps the generator establish a foundational representation before being fine-tuned in an adversarial framework \cite{https://doi.org/10.48550/arxiv.1809.00219}. Following pretraining, the model transitions to a relativistic adversarial setup. A relativistic discriminator $D_{\text{Ra}, C}$ evaluates the probability that real data are more realistic than generated data \cite{https://doi.org/10.48550/arxiv.1809.00219}:
\begin{equation}
    D_{\text{Ra}, C}(a, b) = \sigma \bigl(C(a) - \mathbb{E}_b \bigl[C(b)\bigr]\bigr),
\end{equation}
where $\sigma$ is the sigmoid function, $C(\cdot)$ represents the encoded ($C_{\text{enc}}$) or decoded ($C_{\text{dec}}$) discriminator output, and $\mathbb{E}_b$ denotes the expectation over the batch. The discriminator's loss includes terms for both encoder and decoder outputs:
\begin{equation}
    \begin{aligned}
L_{D,\text{enc}}^{\text{Ra}} &= -\mathbb{E}_{x_r} \Bigl[\log \bigl(D_{\text{Ra}, C_{\text{enc}}}(x_r, x_f)\bigr)\Bigr] \\
&\quad - \mathbb{E}_{x_r} \Bigl[\log \bigl(1 - D_{\text{Ra}, C_{\text{enc}}}(x_f, x_r)\bigr)\Bigr], \\[6pt]
L_{D,\text{dec}}^{\text{Ra}} &= -\mathbb{E}_{x_f} \Bigl[\mathbb{E}_{p} \Bigl[\log \bigl(D_{\text{Ra}, C_{\text{dec}}}(x_r, x_f)\bigr)\Bigr]\Bigr] \\
&\quad - \mathbb{E}_{x_f} \Bigl[\mathbb{E}_{p} \Bigl[\log \bigl(1 - D_{\text{Ra}, C_{\text{dec}}}(x_f, x_r)\bigr)\Bigr]\Bigr].
\end{aligned}
\end{equation}
Using consistency loss terms ensures robust alignment when CutMix augmentation is applied \cite{schonfeld2020u}:
\begin{equation}
\begin{aligned}
    L_{D,\text{cons}} = \mathbb{E}_{x_m} \Bigl[ &\bigl\| C_{\text{dec}} \bigl(\mathrm{mix}(x_r, x_f, M)\bigr) \\
    &- \mathrm{mix} \bigl(C_{\text{dec}}(x_r), C_{\text{dec}}(x_f), M\bigr) \bigr\|^2 \Bigr],
\end{aligned}
\end{equation}
where $\mathrm{mix}(\cdot)$ applies a mask $M$ to merge real and generated data, and $\|\cdot\|^2$ is the $L_2$-norm, penalizing inconsistencies in the discriminator response. Incorporating these components, the overall discriminator loss is expressed as:
\begin{equation}
    L_D = \alpha \bigl(L_{D,\text{enc}}^{\text{Ra}} + L_{D,\text{dec}}^{\text{Ra}}\bigr) + \beta \, L_{D,\text{cons}}.
\end{equation}
The generator's objective function combines both the previously introduced $L_{G,1}$ term and the relativistic adversarial losses:
\begin{equation}
    \begin{aligned}
L_{G,\text{enc}}^{\text{Ra}} &= -\mathbb{E}_{x_r} \Bigl[\log \bigl(1 - D_{\text{Ra}, C_{\text{enc}}}(x_r, x_f)\bigr)\Bigr] \\
&\quad - \mathbb{E}_{x_f} \Bigl[\log \bigl(D_{\text{Ra}, C_{\text{enc}}}(x_f, x_r)\bigr)\Bigr], \\[6pt]
L_{G,\text{dec}}^{\text{Ra}} &= -\mathbb{E}_{x_r} \Bigl[\mathbb{E}_{p} \Bigl[\log \bigl(1 - D_{\text{Ra}, C_{\text{dec}}}(x_r, x_f)\bigr)\Bigr]\Bigr] \\
&\quad - \mathbb{E}_{x_f} \Bigl[\mathbb{E}_{p} \Bigl[\log \bigl(D_{\text{Ra}, C_{\text{dec}}}(x_f, x_r)\bigr)\Bigr]\Bigr], \\[6pt]
L_G &= L_{G,1} + \gamma \,\Bigl(L_{G,\text{enc}}^{\text{Ra}} + L_{G,\text{dec}}^{\text{Ra}}\Bigr).
\end{aligned}
\end{equation}
This multifaceted objective encourages the generator to create high-fidelity outputs that align with ground truth. Subsequently, hyperparameters are tuned on the held-out monthly validation set.

\subsection{Evaluation metrics}\label{sec:eval_metrics}
In this study, the results are evaluated based on three metrics: root-mean-square error (RMSE), correlation ($\rho$), and explained variance (VAR). RMSE quantifies absolute discrepancies between observed $(x_{\text{obs}})$ and predicted $(x_{\text{pred}})$ TWSA values according to
\begin{equation}
    \text{RMSE} = \sqrt{\frac{1}{N} \sum_{i=1}^N \bigl(x_{\text{obs},i} - x_{\text{pred},i}\bigr)^2},
\end{equation}
where $N$ is the number of data points. Lower RMSE indicates closer agreement with observations. In hydrologically dynamic regions, larger RMSE values can partly reflect the larger signal amplitude itself, so RMSE should be interpreted together with correlation and VAR rather than in isolation \cite{Eicker2020}.

The correlation coefficient measures the linear association between observed and predicted series:
\begin{equation}
    \rho = \frac{\mathrm{Cov}\bigl(x_{\text{obs}}, x_{\text{pred}}\bigr)}{\sigma_{\text{obs}}\,\sigma_{\text{pred}}},
\end{equation}
where $\mathrm{Cov}\bigl(x_{\text{obs}}, x_{\text{pred}}\bigr)$ is the covariance between observed and predicted values, and $\sigma_{\text{obs}}, \sigma_{\text{pred}}$ are the respective standard deviations. High correlation implies that the model effectively tracks the timing and phase of TWSA fluctuations, even if magnitude discrepancies exist \cite{Eicker2020}.

VAR evaluates how much observed variability is captured by the predictions:
\begin{equation}
    \text{VAR} = 1 - \frac{\mathrm{Var}\bigl(x_{\text{obs}} - x_{\text{pred}}\bigr)}{\mathrm{Var}\bigl(x_{\text{obs}}\bigr)},
\end{equation}
where $\mathrm{Var}\bigl(x_{\text{obs}} - x_{\text{pred}}\bigr)$ is the variance of residuals, and $\mathrm{Var}\bigl(x_{\text{obs}}\bigr)$ is the variance of observed data. Values near +1 indicate that the model explains virtually all observed variability, whereas negative values suggest that the model introduces noise or fails to represent important physical processes \cite{Eicker2020}.

\section{Results and Discussion}\label{sec:results}
\subsection{Model Evaluation on Monthly Scale}
\begin{figure*}[!tb]
    \centering
    \includegraphics[width=\textwidth]{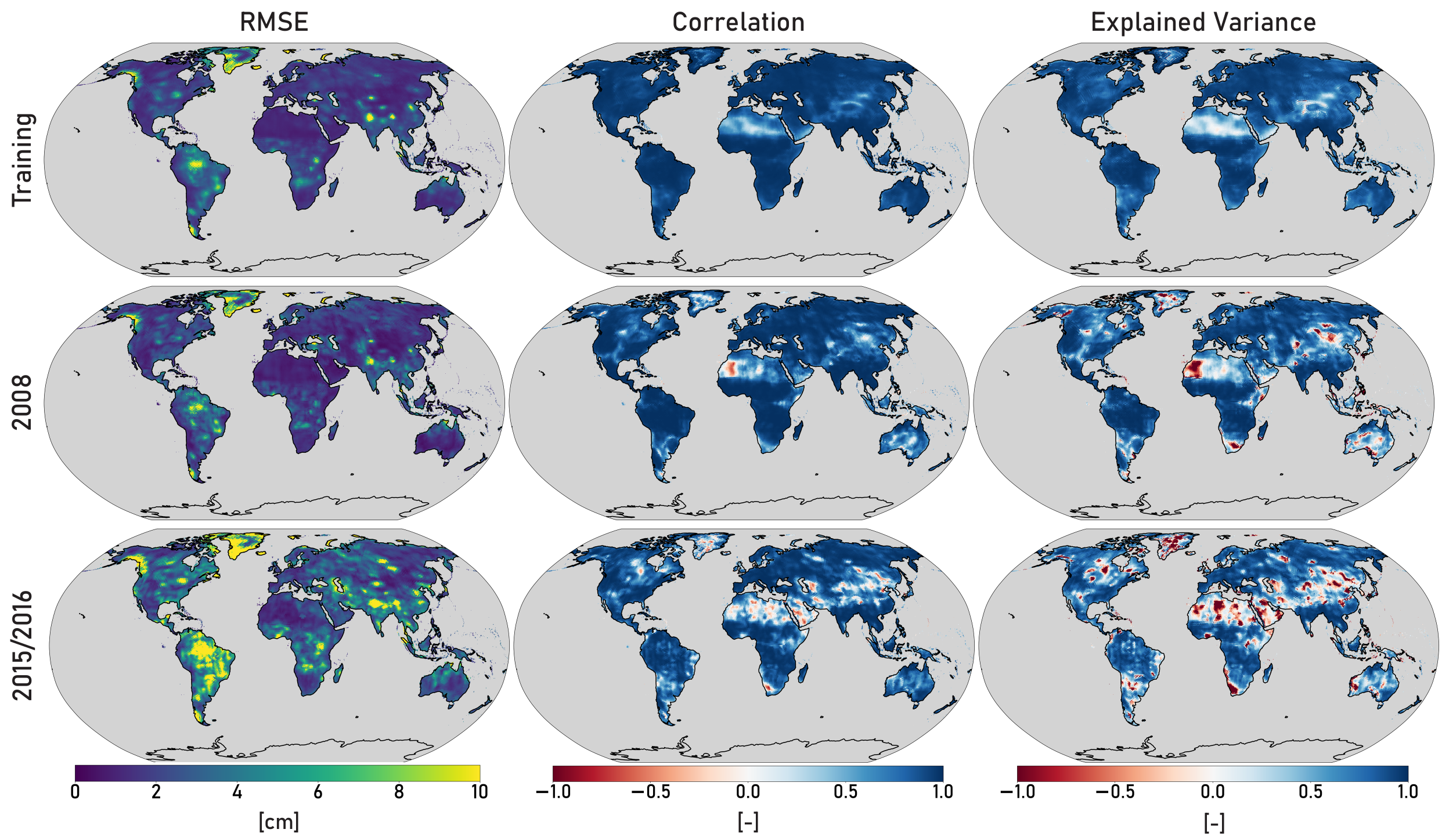}
    \caption{Latitude-weighted global means of RMSE [\si{\centi\meter}], correlation ($\rho$), and explained variance (VAR) comparing predictions to monthly CSR mascon solutions for the training period, 2008 (monthly validation), and 2015/2016 (monthly validation). Lower RMSE and higher $\rho$/VAR indicate better skill. The training and 2008 patterns are broadly consistent, evidencing generalization; 2015--2016 exhibits degraded metrics consistent with late-mission GRACE noise. These maps contextualize the global averages reported in Table~\ref{tab:evaluation_metrics} and motivate adopting 2008 as the primary comparator year in subsequent analyses.}
    \label{fig:evaluation}
\end{figure*}

We first evaluate the performance of \dshift{} at the monthly scale against available ground truth. In this way, its ability in deriving the relationship between inputs and targets, and generalizability to other time span, are assessed. The monthly-scale evaluation is performed by comparing the performance on the monthly training years to performance on held-out monthly validation years (2008, 2015, and 2016), prior to transferring the model to daily inference. Figure~\ref{fig:evaluation} presents global maps of three evaluation metrics (see Section~\ref{sec:eval_metrics}), for the training set (top row), 2008 (monthly validation, middle row), and 2015/2016 (monthly validation, bottom row). Latitude-weighted global average values for each metric are listed in Table~\ref{tab:evaluation_metrics}, facilitating numerical comparisons between the training and validation periods. In the training data, the top-row maps illustrate that most regions have low RMSE (\textless~\SI{4}{\centi\meter} as a typical small-basin value \cite{Landerer2012} in purple or dark blue, 90\% of global area), coupled with high correlation (generally in dark blue, \textgreater~0.6, 94\%) and high VAR (generally in dark blue, \textgreater~0.6, 84\% of global area), resulting in 76\% of global area in overlap. Larger errors remain concentrated in high-variability regions such as the Amazon and Greenland, where the absolute TWSA signal is also large. In such regions, the monthly maps should be interpreted jointly: elevated RMSE does not by itself imply poor agreement when correlation and VAR remain comparatively high.

\begin{table}[!t]
\caption{Latitude-Weighted Global Means of RMSE [\si{\centi\meter}], Correlation ($\rho$), and Explained Variance (VAR) for the Training Set, 2008 (Monthly Validation), and 2015--2016 (Monthly Validation). The 2008 Scores Corroborate Generalization at Monthly Scale; the 2015--2016 Degradation Reflects Known Late-Mission GRACE Quality Issues. These Values Provide the Baseline Against Which Daily-Scale Transfer Performance is Interpreted.\label{tab:evaluation_metrics}}
\centering
\begin{tabular}{lccc}
\toprule
\textbf{Metric}                & \textbf{Training} & \textbf{2008} & \textbf{2015-2016} \\ && \textbf{(Validation)} & \textbf{(Validation)} \\
\midrule
RMSE $\downarrow$ [\si{\centi\meter}]             & 2.5                   & 2.4                       & 4.3                           \\
Correlation $\uparrow$ [-]  & 0.9                   & 0.8                       & 0.7                            \\
Explained Variance $\uparrow$ [-] & 0.8                 & 0.6                       & 0.4                            \\
\bottomrule
\end{tabular}
\end{table}

Evaluation in 2008 reveals a spatial distribution of RMSE and correlation that broadly mirrors the metrics on the training set, suggesting good temporal generalizability of \dshift{}. Some declines in correlation are noticeable in regions such as the Sahara with relatively low RMSE, which are due to generally weak hydrological signals. The VAR map for 2008 remains predominantly positive with slightly increasing amount of negative values than in the training data. This outcome implies that the model captures substantial fractions of TWSA variability across the globe but experiences difficulty in specific locations when validated on a different year. The markedly poorer performance in 2015--2016 primarily reflects the known degradation in GRACE data quality during the mission's terminal phase \cite{Boergens2022, liwiska2020}, which caused a data distribution shift compared to earlier years tahta re used for training the model~\cite{gentner2026deeprec}. Accordingly, 2008 is used as the primary comparator for subsequent analyses. Table~\ref{tab:evaluation_metrics} shows the averaged metrics. The results show that RMSE values of around \SIrange{2.4}{2.5}{\centi\meter} are within the established GRACE uncertainty range (\SIrange{2}{3}{\centi\meter}; \cite{Wahr2006}), confirming that the model's predictions align well with observational uncertainty levels. Correlations exceed 0.8, indicating consistent modeling of temporal variations, specifically without obvious phase shift. The VAR remains above 0.5 in global mean, reflecting strong overall skill in reproducing interannual and seasonal signals. Large areas with active signals, such as the Amazon and Greenland, exhibit greater absolute errors but still show substantial correlation and VAR. It is because that the elevated RMSE values primarily stem from the magnitude of hydrological variations rather than model's deficiency. Conversely, low-signal regions (such as the Sahara or parts of Central Africa) display weaker correlation and negative explained variance, where limited observed variability, fewer ground constraints, or unmodeled processes introduce uncertainty into the predictions.


\subsection{Evaluation of Daily Solution}\label{sec:eval}
\begin{figure*}[!h]
    \centering
    \includegraphics[width=0.8\textwidth]{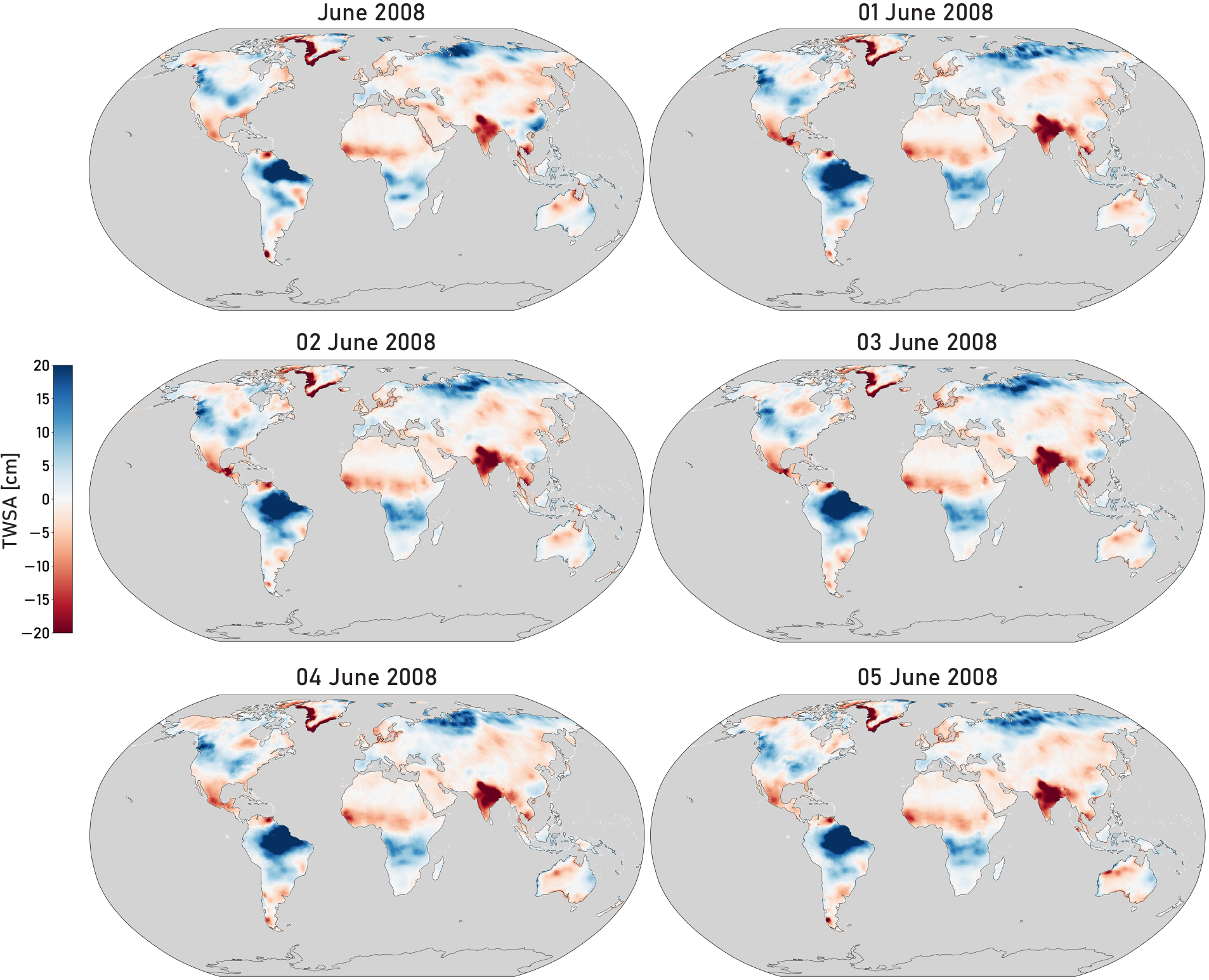}
    \caption{Monthly (June 2008) and daily (1--5 June 2008) TWSA products on a \SI{0.25}{\degree} grid. The monthly map highlights the large-scale anomaly structure, while the consecutive daily maps show the corresponding daily fields during the first five days of June. The figure provides a representative visual comparison of large-scale consistency and short-term daily variability within the monthly validation period.}
    \label{fig:examples}
\end{figure*}
The daily TWSA product generated by \dshift{} represents the central evaluation of the proposed monthly-to-daily transfer framework and should reveal fluctuations at daily scale that are inherently smoothed or averaged out in monthly fields. The lack of reliable ground-truth data at a daily timescale requires an alternative approach to evaluating these higher-frequency datasets. This section evaluates the daily solutions' spatiotemporal variations, visual coherence (including short-term variability) and assesses mass conservation over large basins (area \textgreater~\SI{200000}{\kilo\meter\squared}; HydroSHEDS, Section~\ref{sec:basins}). The \SI{200000}{\kilo\meter\squared} criterion is used to obtain more reliable quantification from GRACE-based observations \cite{Vishwakarma2018, Boergens2022}. Trend and seasonal signals of the TWSA time series are subsequently estimated, followed by a comparison of derived parameters with the monthly CSR mascon reference. To assess spatial resolution gains, a basin-area--dependent double-difference of trend and seasonal errors relative to CSR Monthly is evaluated, with basins larger than about \SI{63000}{\kilo\meter\squared} treated as the limiting resolution of monthly GRACE L3 products \cite{Vishwakarma2018,Gou2024}.

Global maps of daily TWSA retain the principal seasonal and interannual features visible in monthly products, yet they expose additional high-frequency variability (Fig.~\ref{fig:examples}). Highly dynamic regions, such as the Amazon Basin or monsoon-influenced portions of Asia, particularly the Ganges--Brahmaputra region and adjacent coastal catchments, exhibit transient fluctuations that are less pronounced in monthly averages. The large-scale distribution of positive (wet) and negative (dry) anomalies remains broadly consistent with monthly fields. These observations confirm that daily solutions capture overarching hydrological patterns while with finer temporal resolution.

\begin{figure*}[!t]
    \centering
    \includegraphics[width=0.8\textwidth]{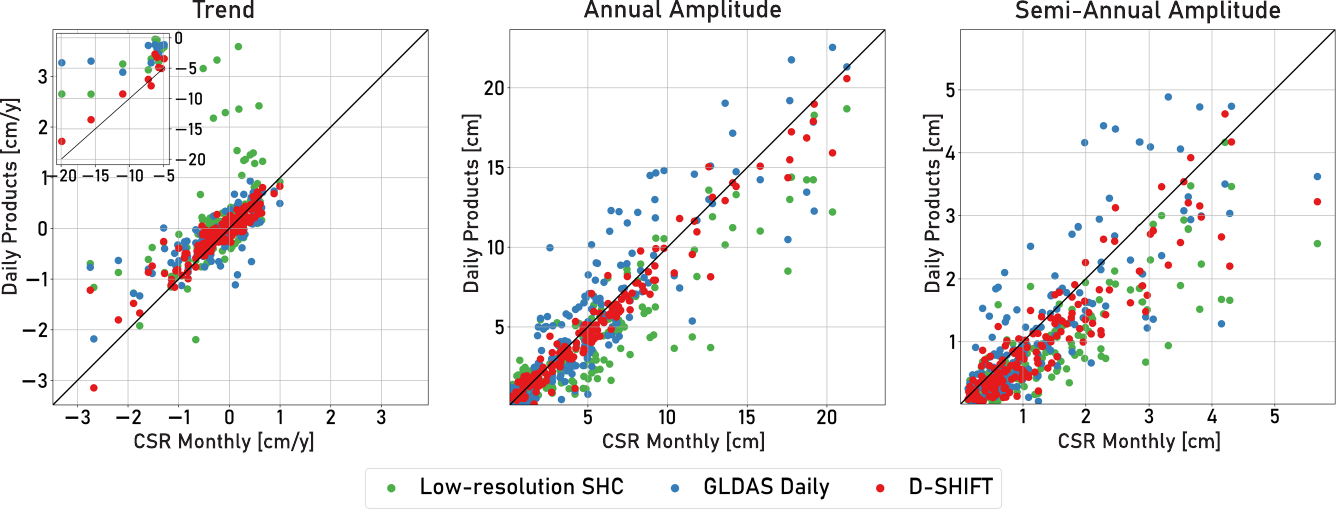}
    \caption{Scatter comparisons of trend [\si{\centi\meter\per\year}], annual, and semi-annual amplitudes [\si{\centi\meter}] estimated for basins with area \textgreater~\SI{200000}{\kilo\meter\squared}. Each point corresponds to one basin, plotted as $S_b^\text{dataset}$ versus $S_b^\text{CSR}$; the 1:1 line denotes perfect agreement. Proximity to the line indicates how closely each daily product reproduces CSR Monthly's low-frequency behavior at basin scale. Systematic offsets from the line reveal under/over-estimation of seasonal energy or trends, and increased scatter suggests leakage or smoothing effects that vary by region and product.}
    \label{fig:coefficients}
\end{figure*}

We further estimated trends, annual and semi-annual signals from both daily and monthly TWSA within each large basin as follows:
\begin{equation}
\begin{aligned}
    \text{TWSA}(t) = {} & a + bt + c \cos\!\Bigl(\tfrac{2\pi t}{T}\Bigr) + d \sin\!\Bigl(\tfrac{2\pi t}{T}\Bigr) \\
    & + e \cos\!\Bigl(\tfrac{4\pi t}{T}\Bigr) + f \sin\!\Bigl(\tfrac{4\pi t}{T}\Bigr),
\end{aligned}
\end{equation}
where \(a\) and \(b\) represent an offset and a linear trend, while $A = \sqrt{c^2 + d^2}$ and $B = \sqrt{e^2 + f^2}$ correspond to the amplitudes of annual and semiannual cycles, respectively. Estimating these coefficients allow us to compare daily basin signals against the CSR Monthly reference, understanding whether daily TWSA follows the long-term changes and seasonality contained in monthly solutions (see Figure~\ref{fig:coefficients}). Our \dshift{} approach systematically show better agreements to the CSR mascon reference across all three parameters, implying a higher fidelity in reproducing low-frequency TWSA variations than the low-resolution daily solution and model simulations. This implies the higher effective spatial resolution of \dshift{}-generated daily solution which reduces inter-basin leakage errors.

\begin{figure*}[!t]
    \centering
    \includegraphics[width=0.8\textwidth]{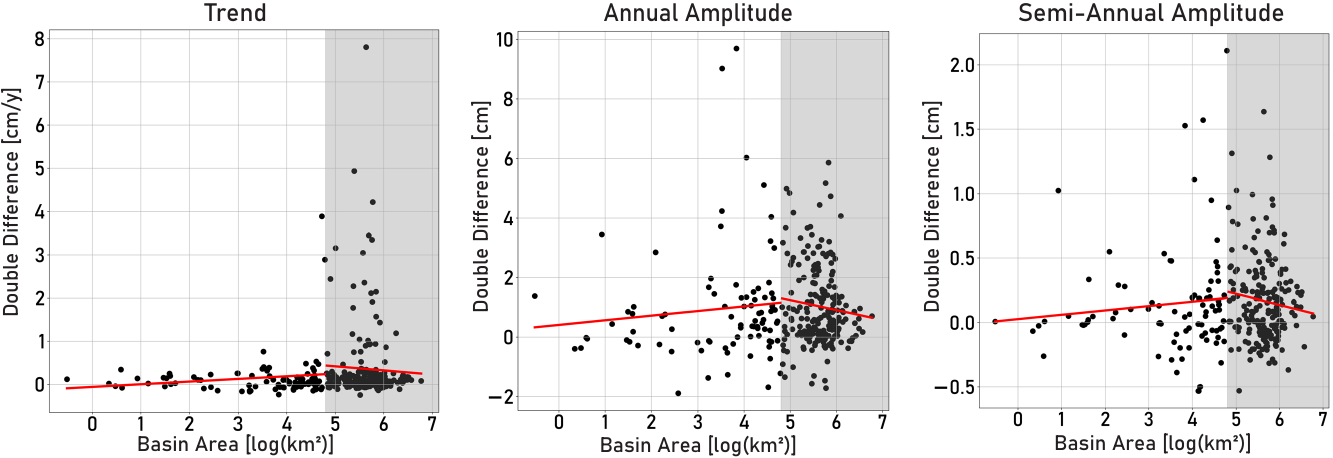}
    \caption{Double-difference of basin-wise errors versus basin area ($\log_{10}$ scale) for (left) trend, (middle) annual amplitude, and (right) semi-annual amplitude. For each basin $b$, the metric $\Delta\epsilon_b=|S_b^\text{ref}-S_b^\text{comp}|-|S_b^\text{ref}-S_b^\text{pred}|$ compares the low-resolution SHC (``comp'') and our daily high-resolution solution (``pred'') against the CSR monthly reference (``ref''); positive values indicate that our daily solution is closer to the reference. The gray band marks basins \textgreater~\SI{63000}{\kilo\meter\squared}, a commonly used lower bound for reliable GRACE signal recovery. Red lines show least-squares fits computed separately below and above the \SI{63000}{\kilo\meter\squared} threshold, highlighting the area dependence: gains are largest for moderate-size basins, attenuate for large basins, and are variable for very small basins.}
    \label{fig:ddiff}
\end{figure*}

As spatial resolution affects leakage errors and small-basin noise, we evaluate the performance of \dshift{} as a function of basin area. Following common practice, \SI{63000}{\kilo\meter\squared} is used as a practical lower bound for reliable GRACE signal recovery; hence, we emphasize basins above this threshold \cite{Vishwakarma2018}. No upper bound is imposed, since for sufficiently large basins both low- and high-resolution products converge. Reference ``closeness'' is measured via a double-difference:
\begin{equation}
    \Delta\epsilon_b=|S_b^\text{ref}-S_b^\text{comp}|-|S_b^\text{ref}-S_b^\text{pred}|,
\end{equation}
where $S_b$ denotes the basin-wise trend or seasonal amplitude for basin $b$, ``ref'' refers to the CSR monthly product, ``comp'' refers to the low-resolution SHC, and ``pred'' refers to our daily high-resolution solution. Thus, $\Delta\epsilon_b>0$ indicates that our daily product is closer to the reference than the low-resolution SHC daily product for that basin and signal component. As seen in Figure~\ref{fig:ddiff}, across basins, $\Delta\epsilon_b$ exhibits a clear area dependence. Improvements ($\Delta\epsilon_b>0$) are most frequent and of largest magnitude for basins in the moderate-size range (\textgreater~\SI{63000}{\kilo\meter\squared} and below a few \SI{e5}{\kilo\meter\squared}), consistent with the expectation that higher effective resolution reduces leakage and improves the estimation of trend and seasonal amplitudes in relatively small basins. For very small basins (\textless~\SI{63000}{\kilo\meter\squared}), results are more variable, reflecting limited ability of GRACE to observe hydrological signals at that scale. For large basins, the advantage becomes less prominent, as the low-resolution daily solution can also resolve the signals in these basins accurately, causing $\Delta\epsilon_b$ to approach zero.

\begin{figure*}[!t]
    \centering
    \includegraphics[width=0.8\textwidth]{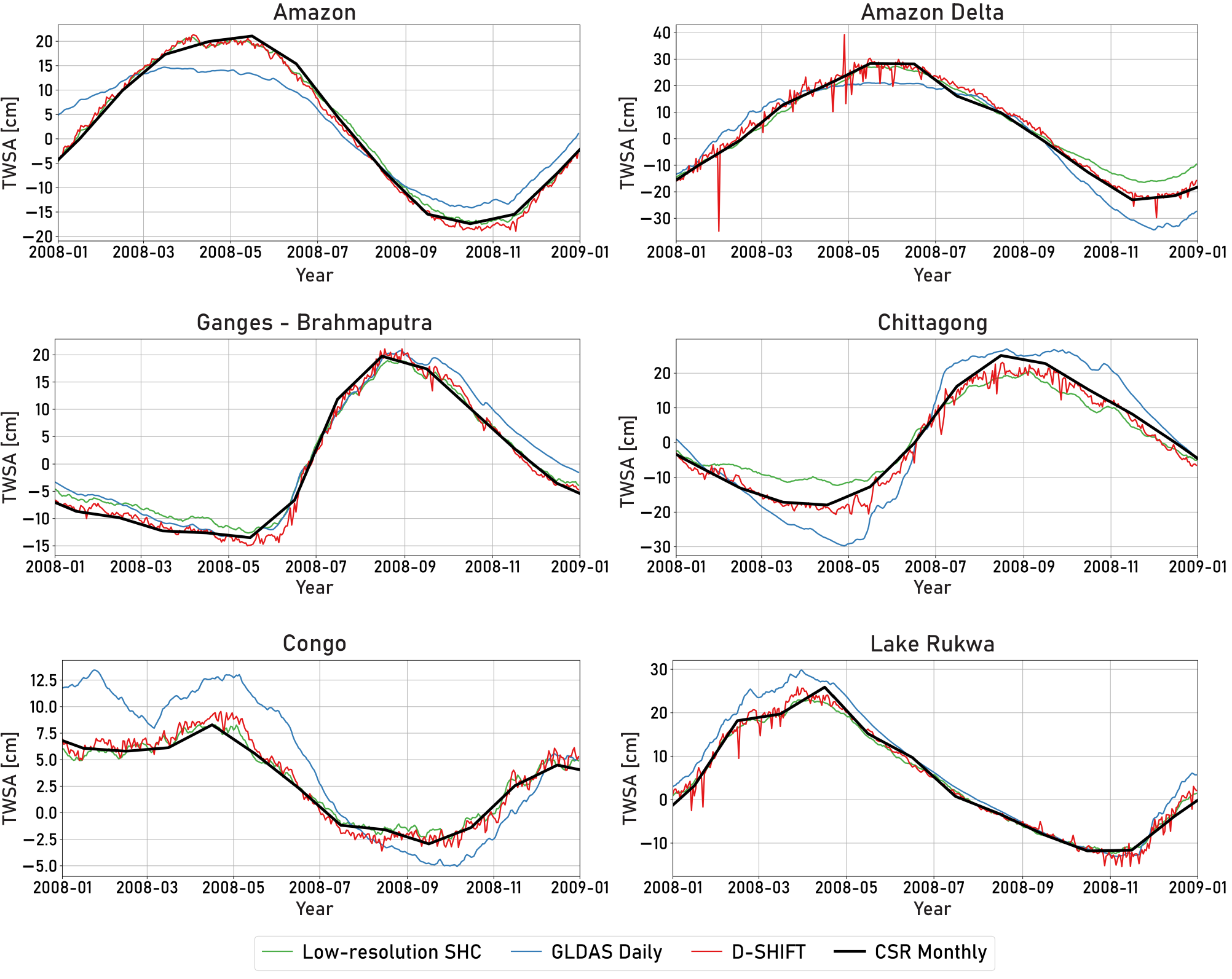}
    \caption{Area-mean TWSA time series during 2008 for six representative basins (Amazon, Amazon Delta, Ganges--Brahmaputra, Chittagong, Congo, and Lake Rukwa) from CSR Monthly and the daily products. The one-year window is shown for readability and because 2008 is the primary monthly validation year. It supports visual comparison of seasonal phase, seasonal amplitude, and daily-scale noise; long-term trends are quantified from the full-period harmonic fits in Table~\ref{tab:basin_metrics}.}
    \label{fig:timeseries}
\end{figure*}

\begin{table*}[!t]
\caption{Trends [\si{\centi\meter\per\year}], Annual, and Semi-Annual Amplitudes [\si{\centi\meter}] Estimated from Harmonic Fits to Area-Mean TWSA for Six Basins (Three Large, Three Nearby Smaller). CSR Monthly Serves as the Reference; Boldface Highlights the Closest Daily Value to CSR for Each Metric (Smallest Absolute Difference). The Table Complements Figure~\ref{fig:timeseries}, Quantifying How Well Each Product Captures Large-Scale Behavior (Trends) and Seasonal Magnitudes, and Illustrating the Area-Dependent Performance Differences Emphasized by the Double-Difference Analysis.\label{tab:basin_metrics}}
\centering
\begin{tabular}{lrlccc}
\toprule
\textbf{Basin}            & \textbf{Area} & \textbf{Dataset}               & \textbf{Trend} & \textbf{Annual} & \textbf{Semiannual} \\
                          & \textbf{[\si{\kilo\meter\squared}]} &                                & \textbf{[\si{\centi\meter\per\year}]} & \textbf{Amplitude} & \textbf{Amplitude} \\
                          &&                                &                & \textbf{[\si{\centi\meter}]}    & \textbf{[\si{\centi\meter}]}      \\
\midrule
Amazon & 5'912'923   & low-resolution SHC      & 0.3          & 18.3           & 0.8               \\
                          && GLDAS Daily                   & -0.4         & 12.3           & 0.7               \\
                          && \dshift{}                     & \textbf{0.2} & \textbf{19.0}           & \textbf{0.9}      \\
                          && \textbf{CSR Monthly}          & \textbf{0.2} & \textbf{19.2}  & \textbf{1.0}      \\\midrule
Congo & 3'705'222    & low-resolution SHC      & 0.3         & 4.3            & \textbf{1.6}               \\
                          && GLDAS Daily                   & -1.1         & 5.1            & 1.8               \\
                          && \dshift{}                     & \textbf{0.1} & \textbf{4.9}   & 1.8               \\
                          && \textbf{CSR Monthly}          & \textbf{0.1} & \textbf{4.7}   & \textbf{1.6}      \\\midrule
Ganges & 1'584'745
                          & low-resolution SHC     & -0.6         & 11.9           & \textbf{4.2}      \\
--Brahmaputra              && GLDAS Daily                   & -0.8         & \textbf{12.7}           & 3.5               \\
                          && \dshift{}                     & \textbf{-0.9} & 13.1          & 4.6               \\
                          && \textbf{CSR Monthly}          & \textbf{-1.6} & \textbf{12.8}  & \textbf{4.2}      \\\midrule
Chittagong & 96'502   & low-resolution SHC      & \textbf{-0.2}          & 15.2           & 2.6               \\
                          && GLDAS Daily                   & \textbf{-0.2}         & 28.9          & 4.8              \\
                          && \dshift{}                     & -0.3 & \textbf{20.1}           & \textbf{3.0}      \\
                          && \textbf{CSR Monthly}          & \textbf{-0.1} & \textbf{21.4}  & \textbf{3.4}      \\\midrule
Lake Rukwa & 79'611   & low-resolution SHC      & \textbf{0.2}          & 14.1           & 1.4               \\
                          && GLDAS Daily                   & 0.4         & 15.8          & \textbf{2.3}               \\
                          && \dshift{}                     & \textbf{0.2} & \textbf{14.9}           & 1.9      \\
                          && \textbf{CSR Monthly}          & \textbf{0.2} & \textbf{15.1}  & \textbf{2.3}      \\\midrule
Amazon Delta & 37'354   & low-resolution SHC      & 0.4          & 19.6          & 1.1               \\
                          && GLDAS Daily                   & -1.2        & \textbf{25.8}           & \textbf{3.8}              \\
                          && \dshift{}                     & \textbf{0.2} & 22.9          & 1.6      \\
                          && \textbf{CSR Monthly}          & \textbf{0.2} & \textbf{24.4}  & \textbf{3.7}      \\
\bottomrule
\end{tabular}
\end{table*}

Table~\ref{tab:basin_metrics} summarizes full-period harmonic fits for three large basins (Amazon, Congo, and Ganges--Brahmaputra) and three smaller basins in hydrologically complex regions (Amazon Delta, Lake Rukwa, and Chittagong). Figure~\ref{fig:timeseries} shows the corresponding basin-mean time series for 2008 to highlight seasonal phases, amplitudes, and daily variability visually. Across the large basins, \dshift{} is generally close to CSR Monthly for the fitted low-frequency parameters. In the Amazon Basin, \dshift{} follows the CSR trend and annual amplitude closely (\SI{0.2}{\centi\meter\per\year} and \SI{19.0}{\centi\meter}, compared with \SI{0.2}{\centi\meter\per\year} and \SI{19.2}{\centi\meter} for CSR). In the Congo Basin, where the seasonal variation is weaker, \dshift{} remains close to CSR for both trend and annual amplitude, whereas GLDAS Daily shows a stronger negative trend implying the limitation of land surface model simulations. In the Ganges--Brahmaputra Basin, \dshift{} gives the closest trend estimates compared to CSR, while GLDAS Daily and low-resolution SHC are slightly closer for the annual and semiannual amplitudes, respectively. Thus, the advantage of \dshift{} is not uniform across all metrics, but it is most consistent for retaining CSR-like trends and seasonal magnitudes in basins where leakage and smoothing are important.

The smaller basins illustrate the area-dependent behavior seen in Figure~\ref{fig:ddiff}. For Chittagong, \dshift{} is closest to CSR for the annual and semiannual amplitudes, although the low-resolution SHC and GLDAS trends are slightly closer. For Lake Rukwa, \dshift{} matches the CSR trend and is closest for the annual amplitude, while GLDAS Daily is closest for the semiannual term. In the Amazon Delta, \dshift{} matches the CSR trend, but GLDAS Daily is closer for the fitted annual and semiannual amplitudes. The 2008 time series support these parameter-based comparisons qualitatively: \dshift{} generally follows the seasonal phase of CSR Monthly, but it also contains visibly higher high-frequency variability. This additional variability is expected for a daily inferred product and should be interpreted together with the harmonic-fit metrics rather than as a standalone proof of improved performance.

\begin{figure*}[!t]
    \centering
    \includegraphics[width=0.8\textwidth]{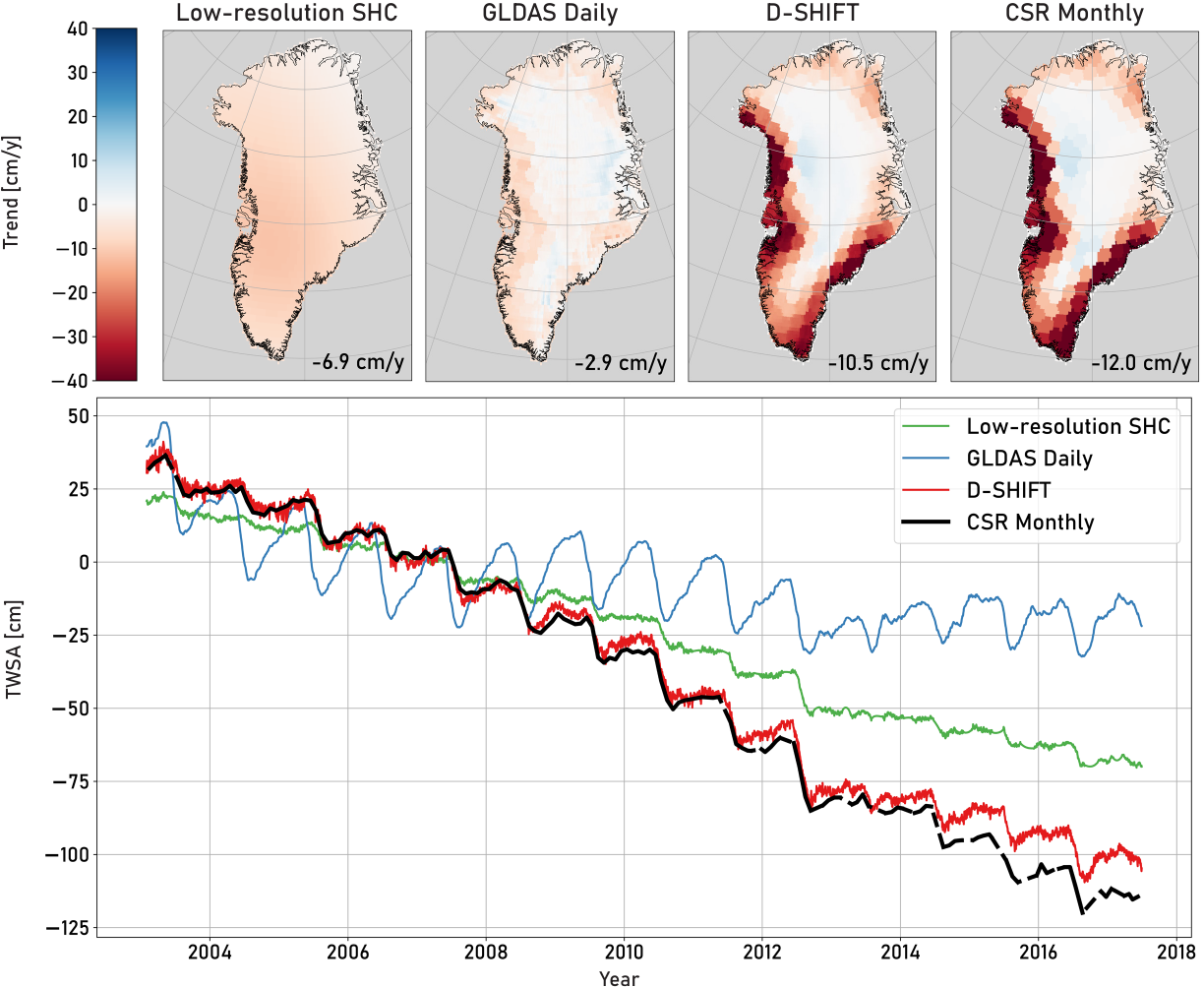}
    \caption{Top: pixel-wise trend maps [\si{\centi\meter\per\year}] for daily products alongside CSR Monthly, emphasizing coastal mass-loss zones where land--ocean separation is most challenging. Bottom: Greenland-mean TWSA time series demonstrating how daily products track the cumulative melt signal. \dshift{} shows sharper coastal delineation and closer trend magnitude agreement with CSR, whereas other daily products exhibit deviations in amplitude. Accurate coastline representation is essential to quantify ice-sheet mass change without oceanic contamination.}
    \label{fig:greenland}
\end{figure*}

Unlike the hydrological basins considered above, Greenland represents a cryospheric case in which the dominant signal is caused by continuous mass loss concentrated in narrow coastal outlet regions (the ablation zone) \cite{2019, Wouters2008}, while the signal leakages at the land-ocean boundaries are prominent. The limiting spatial resolution of daily SHC-based solution causes serve signal leakages from the Greenland ice sheet (GrIS) into the ocean. resulting in underestimate of GrIS mass losses \cite{Schrama2011}. We therefore use Greenland as a targeted case study to evaluate whether\dshift{} improves the situation by re-concentrating the large, signals resolved by monthly reference back to the Greenland. Figure~\ref{fig:greenland} shows that \dshift{} preserves a cleaner boundary along the coastline, reflecting CSR Monthly's pattern of negative trends more accurately than certain other daily products. The multi-year Greenland time series confirms the performance of \dshift{} that it tracks the monthly record's cumulative mass loss while low-resolution daily SHC and GLDAS simulations deviate in both amplitude and temporal phases.

\section{Conclusions and Outlook}\label{sec:conclusions}

In this study, we proposed \dshift{} (Daily Spatial High-Resolution Inference via Feature Transformation), a deep-learning framework for generating daily high-resolution TWSA fields from daily SHC-based solutions by transferring the relationship learned on the monthly scale. The model is based on a modified nESRGAN+ generator and a U-Net discriminator with basin-aware CutMix augmentation. Training was performed in the monthly domain using truncated ITSG monthly products with auxiliary data as inputs while targeting on CSR monthly mascon. It is then applied to daily ITSG inputs without daily retraining, as daily ground truth is inaccessible.

Compared to monthly CSR mascon, \dshift{} reaches an RMSE of \SI{2.3}{\centi\meter}. This confirms that the model can capture the relationship between mascon-like high-resolution spatial details and low-resolution SHC inputs, which is the basis for the daily inference. The daily analyses then evaluate different aspects of the inferred product. The 2008 basin time series test whether the daily product follows the seasonal behavior of CSR Monthly while providing sub-monthly variability. These examples show that \dshift{} often follows the monthly reference more closely than low-resolution SHC, but also contains stronger high-frequency variability. The harmonic basin analysis evaluates the full-period trend, annual amplitude, and semi-annual amplitude of basin-mean TWSA. It shows that \dshift{} improves several trend and seasonal-amplitude estimates compared with low-resolution SHC, but not for every basin and every fitted parameter. The basin-area double-difference analysis directly tests where \dshift{} reduces the discrepancy to CSR Monthly relative to the low-resolution daily product. This analysis supports the main spatial-resolution claim: the benefit is strongest for spatially localized signals where smoothing and leakage are more limiting. The Greenland case study evaluates the method in the region, where land--ocean leakage and narrow coastal mass-loss signals are particularly important. In the trend maps, \dshift{} better reproduces the coastal mass-loss pattern of CSR Monthly than the other daily products. In the basin-mean time series, it also follows mass loss signal  estimated from monthly reference more clearly. The basin-mean trend is \SI{-10.5}{\centi\meter\per\year} for \dshift{}, close to the CSR Monthly value of \SI{-12.0}{\centi\meter\per\year}.

Overall, the results show that \dshift{} can transfer monthly mascon-like spatial information to daily GRACE-based inputs and produce daily TWSA fields with improved spatial detail compared with low-resolution SHC. It provides a new venue for generating high-frequency TWSA product with potential contribution to various downstream applications, such as flash hydrological extremes monitoring. The main limitation is the increased high-frequency noise in the generated daily time series. Some isolated daily excursions are not clearly supported by the monthly reference and should therefore be interpreted cautiously. Future work should focus on improving temporal consistency, for example through explicit temporal constraints or sequence-based model components, and on validating selected daily signals with independent hydrological, cryospheric, or in-situ observations.


\section*{Data Availability}
The enhanced daily product, all the codes, and other intermediate datasets are available from the corresponding author upon a reasonable request.

\section*{Author Contributions}
\textbf{Andreas Dombos}: Methodology, Software, Validation, Formal analysis, Investigation, Data Curation, Visualization, Writing -- original draft, Writing -- review \& editing.
\textbf{Junyang Gou}: Conceptualization, Supervision, Methodology, Investigation, Data Curation, Writing -- original draft, Writing -- review \& editing.
\textbf{Benedikt Soja}: Conceptualization, Supervision, Writing -- review \& editing.

\section*{Acknowledgment}
A part of this study has been performed during Junyang Gou's scientific visit at Bodner lab, MIT.

\bibliographystyle{IEEEtran}
\bibliography{reference}

@article{gou2025DSOBP,
  title={Downscaling GRACE-derived ocean bottom pressure anomalies using self-supervised data fusion},
  author={Gou, Junyang and B{\"o}rger, Lara and Schindelegger, Michael and Soja, Benedikt},
  journal={Journal of Geodesy},
  volume={99},
  number={2},
  pages={19},
  year={2025},
  publisher={Springer},
  doi={https://doi.org/10.1007/s00190-025-01943-9}
}

@article{gentner2026deeprec,
  title={{DeepRec: Global terrestrial water storage reconstruction since 1941 using spatiotemporal-aware deep learning model}},
  author={Gentner, Luis Q and Gou, Junyang and Tourian, Mohammad J and B{\"o}rger, Lara and Sneeuw, Nico and Soja, Benedikt},
  journal={Journal of Geophysical Research: Machine Learning and Computation},
  volume={3},
  number={1},
  pages={e2025JH000889},
  year={2026},
  publisher={Wiley Online Library},
  doi={https://doi.org/10.1029/2025JH000889}
}

@article{tapley2004GracePrinciple,
  title={{The gravity recovery and climate experiment: Mission overview and early results}},
  author={Tapley, Byron D and Bettadpur, S and Watkins, Mo and Reigber, Ch},
  journal={Geophysical Research Letters},
  volume={31},
  number={9},
  year={2004},
  publisher={Wiley Online Library},
  doi={10.1029/2004GL019920}
}

@article{wahr2004GracePrinciple,
  title={{Time-variable gravity from GRACE: First results}},
  author={Wahr, John and Swenson, Sean and Zlotnicki, Victor and Velicogna, Isabella},
  journal={Geophysical Research Letters},
  volume={31},
  number={11},
  year={2004},
  publisher={Wiley Online Library},
  doi={10.1029/2004GL019779}
}

@article{landerer2020GRACE-FO,
  title={{Extending the global mass change data record: GRACE Follow-On instrument and science data performance}},
  author={Landerer, Felix W and Flechtner, Frank M and Save, Himanshu and Webb, Frank H and Bandikova, Tamara and Bertiger, William I and Bettadpur, Srinivas V and Byun, Sung Hun and Dahle, Christoph and Dobslaw, Henryk and others},
  journal={Geophysical Research Letters},
  volume={47},
  number={12},
  pages={e2020GL088306},
  year={2020},
  publisher={Wiley Online Library},
  doi={10.1029/2020GL088306}
}

@article{rodell2018EmergingTrend,
  title={{Emerging trends in global freshwater availability}},
  author={Rodell, Matthew and Famiglietti, Jay S and Wiese, David N and Reager, JT and Beaudoing, Hiroko K and Landerer, Felix W and Lo, M-H},
  journal={Nature},
  volume={557},
  number={7707},
  pages={651--659},
  year={2018},
  publisher={Nature Publishing Group},
  doi={https://doi.org/10.1038/s41586-018-0123-1}
}

@article{rodell2023water,
  title={{Water cycle science enabled by the GRACE and GRACE-FO satellite missions}},
  author={Rodell, Matthew and Reager, John T},
  journal={Nature Water},
  volume={1},
  number={1},
  pages={47--59},
  year={2023},
  publisher={Nature Publishing Group UK London},
  doi={https://doi.org/10.1038/s44221-022-00005-0}
}

@article{humphrey2023Review,
  title={{Using satellite-based terrestrial water storage data: a review}},
  author={Humphrey, Vincent and Rodell, Matthew and Eicker, Annette},
  journal={Surveys in Geophysics},
  volume={44},
  number={5},
  pages={1489--1517},
  year={2023},
  publisher={Springer},
  doi={https://doi.org/10.1007/s10712-022-09754-9}
}

@article{vishwakarma2018GRACEResolution,
  title={{What is the spatial resolution of GRACE satellite products for hydrology?}},
  author={Vishwakarma, Bramha Dutt and Devaraju, Balaji and Sneeuw, Nico},
  journal={Remote Sensing},
  volume={10},
  number={6},
  pages={852},
  year={2018},
  publisher={MDPI},
  doi={https://doi.org/10.3390/rs10060852}
}

@article{goodfellow2020GAN,
  title={Generative adversarial networks},
  author={Goodfellow, Ian and Pouget-Abadie, Jean and Mirza, Mehdi and Xu, Bing and Warde-Farley, David and Ozair, Sherjil and Courville, Aaron and Bengio, Yoshua},
  journal={Communications of the ACM},
  volume={63},
  number={11},
  pages={139--144},
  year={2020},
  publisher={ACM New York, NY, USA},
  doi={https://doi.org/10.1145/3422622}
}

@article{Chen2022,
  title = {Applications and Challenges of GRACE and GRACE Follow-On Satellite Gravimetry},
  volume = {43},
  ISSN = {1573-0956},
  url = {http://dx.doi.org/10.1007/s10712-021-09685-x},
  DOI = {10.1007/s10712-021-09685-x},
  number = {1},
  journal = {Surveys in Geophysics},
  publisher = {Springer Science and Business Media LLC},
  author = {Chen,  Jianli and Cazenave,  Anny and Dahle,  Christoph and Llovel,  William and Panet,  Isabelle and Pfeffer,  Julia and Moreira,  Lorena},
  year = {2022},
  month = jan,
  pages = {305–345}
}

@Article{hess-22-2867-2018,
AUTHOR = {Gouweleeuw, B. T. and Kvas, A. and Gruber, C. and Gain, A. K. and Mayer-G\"urr, T. and Flechtner, F. and G\"untner, A.},
TITLE = {Daily GRACE gravity field solutions track major flood events in the
Ganges--Brahmaputra Delta},
JOURNAL = {Hydrology and Earth System Sciences},
VOLUME = {22},
YEAR = {2018},
NUMBER = {5},
PAGES = {2867--2880},
URL = {https://hess.copernicus.org/articles/22/2867/2018/},
DOI = {10.5194/hess-22-2867-2018}
}

@article{https://doi.org/10.1029/2019JB017415,
author = {Kvas, Andreas and Behzadpour, Saniya and Ellmer, Matthias and Klinger, Beate and Strasser, Sebastian and Zehentner, Norbert and Mayer-Gürr, Torsten},
title = {ITSG-Grace2018: Overview and Evaluation of a New GRACE-Only Gravity Field Time Series},
journal = {Journal of Geophysical Research: Solid Earth},
volume = {124},
number = {8},
pages = {9332-9344},
doi = {https://doi.org/10.1029/2019JB017415},
url = {https://agupubs.onlinelibrary.wiley.com/doi/abs/10.1029/2019JB017415},
eprint = {https://agupubs.onlinelibrary.wiley.com/doi/pdf/10.1029/2019JB017415},
year = {2019}
}

@misc{https://doi.org/10.5880/icgem.2018.003,
  doi = {10.5880/ICGEM.2018.003},
  url = {https://dataservices.gfz-potsdam.de/icgem/showshort.php?id=escidoc:3600910},
  author = {Mayer-G\"{u}rr,  Torsten and Behzadpur,  Saniya and Ellmer,  Matthias and Kvas,  Andreas and Klinger,  Beate and Strasser,  Sebastian and Zehentner,  Norbert},
  language = {en},
  title = {ITSG-Grace2018 - Monthly,  Daily and Static Gravity Field Solutions from GRACE},
  publisher = {GFZ Data Services},
  year = {2018},
  copyright = {CC BY 4.0}
}

@article{KURTENBACH201239,
title = {Improved daily GRACE gravity field solutions using a Kalman smoother},
journal = {Journal of Geodynamics},
volume = {59-60},
pages = {39-48},
year = {2012},
note = {Mass Transport and Mass Distribution in the System Earth},
issn = {0264-3707},
doi = {https://doi.org/10.1016/j.jog.2012.02.006},
url = {https://www.sciencedirect.com/science/article/pii/S0264370712000385},
author = {E. Kurtenbach and A. Eicker and T. Mayer-Gürr and M. Holschneider and M. Hayn and M. Fuhrmann and J. Kusche}
}

@article{Save2016,
  title = {High‐resolution CSR GRACE RL05 mascons},
  volume = {121},
  ISSN = {2169-9356},
  url = {http://dx.doi.org/10.1002/2016JB013007},
  DOI = {10.1002/2016jb013007},
  number = {10},
  journal = {Journal of Geophysical Research: Solid Earth},
  publisher = {American Geophysical Union (AGU)},
  author = {Save,  Himanshu and Bettadpur,  Srinivas and Tapley,  Byron D.},
  year = {2016},
  month = oct,
  pages = {7547–7569}
}

@article{save2020csr,
  title={Csr grace and grace-fo rl06 mascon solutions v02},
  author={Save, Himanshu},
  journal={Mascon Solut},
  volume={12},
  pages={24},
  year={2020}
}

@misc{https://doi.org/10.5067/txbmlx370xx8,
  doi = {10.5067/TXBMLX370XX8},
  url = {https://disc.gsfc.nasa.gov/datacollection/GLDAS_CLSM025_DA1_D_2.2.html},
  author = {Li,  Bailing and Beaudoing,  H. and Rodell,  M. and {NASA/GSFC/HSL}},
  title = {GLDAS Catchment Land Surface Model L4 daily 0.25 x 0.25 degree GRACE-DA1,  Version 2.2},
  publisher = {NASA Goddard Earth Sciences Data and Information Services Center},
  year = {2020}
}

@article{Li2019,
  title = {Global GRACE Data Assimilation for Groundwater and Drought Monitoring: Advances and Challenges},
  volume = {55},
  ISSN = {1944-7973},
  url = {http://dx.doi.org/10.1029/2018wr024618},
  DOI = {10.1029/2018wr024618},
  number = {9},
  journal = {Water Resources Research},
  publisher = {American Geophysical Union (AGU)},
  author = {Li,  Bailing and Rodell,  Matthew and Kumar,  Sujay and Beaudoing,  Hiroko Kato and Getirana,  Augusto and Zaitchik,  Benjamin F. and de Goncalves,  Luis Gustavo and Cossetin,  Camila and Bhanja,  Soumendra and Mukherjee,  Abhijit and Tian,  Siyuan and Tangdamrongsub,  Natthachet and Long,  Di and Nanteza,  Jamiat and Lee,  Jejung and Policelli,  Frederick and Goni,  Ibrahim B. and Daira,  Djoret and Bila,  Mohammed and de Lannoy,  Gabriëlle and Mocko,  David and Steele‐Dunne,  Susan C. and Save,  Himanshu and Bettadpur,  Srinivas},
  year = {2019},
  month = sep,
  pages = {7564–7586}
}

@misc{https://doi.org/10.24381/cds.adbb2d47,
  doi = {10.24381/CDS.ADBB2D47},
  url = {https://cds.climate.copernicus.eu/doi/10.24381/cds.adbb2d47},
  author = {{C3S}},
  title = {ERA5 hourly data on single levels from 1940 to present},
  publisher = {Copernicus Climate Change Service (C3S) Climate Data Store (CDS)},
  year = {2018}
}

@article{Gou2024,
  title = {Global high-resolution total water storage anomalies from self-supervised data assimilation using deep learning algorithms},
  volume = {2},
  ISSN = {2731-6084},
  url = {http://dx.doi.org/10.1038/s44221-024-00194-w},
  DOI = {10.1038/s44221-024-00194-w},
  number = {2},
  journal = {Nature Water},
  publisher = {Springer Science and Business Media LLC},
  author = {Gou,  Junyang and Soja,  Benedikt},
  year = {2024},
  month = feb,
  pages = {139–150}
}

@article{Lehner2013,
  title = {Global river hydrography and network routing: baseline data and new approaches to study the world’s large river systems},
  volume = {27},
  ISSN = {1099-1085},
  url = {http://dx.doi.org/10.1002/hyp.9740},
  DOI = {10.1002/hyp.9740},
  number = {15},
  journal = {Hydrological Processes},
  publisher = {Wiley},
  author = {Lehner,  Bernhard and Grill,  G\"{u}nther},
  year = {2013},
  month = apr,
  pages = {2171–2186}
}

@inbook{Garca2014,
  title = {Data Preparation Basic Models},
  ISBN = {9783319102474},
  ISSN = {1868-4408},
  url = {http://dx.doi.org/10.1007/978-3-319-10247-4_3},
  DOI = {10.1007/978-3-319-10247-4_3},
  booktitle = {Data Preprocessing in Data Mining},
  publisher = {Springer International Publishing},
  author = {García,  Salvador and Luengo,  Julián and Herrera,  Francisco},
  year = {2014},
  month = aug,
  pages = {39–57}
}

@book{Goodfellow-et-al-2016,
    title={Deep Learning},
    author={Ian Goodfellow and Yoshua Bengio and Aaron Courville},
    publisher={MIT Press},
    note={\url{http://www.deeplearningbook.org}},
    year={2016}
}

@article{goodfellow2014generative,
  title={Generative adversarial nets},
  author={Goodfellow, Ian and Pouget-Abadie, Jean and Mirza, Mehdi and Xu, Bing and Warde-Farley, David and Ozair, Sherjil and Courville, Aaron and Bengio, Yoshua},
  journal={Advances in neural information processing systems},
  volume={27},
  year={2014}
}

@inproceedings{rakotonirina2020esrgan+,
  title={ESRGAN+: Further improving enhanced super-resolution generative adversarial network},
  author={Rakotonirina, Nathana{\"e}l Carraz and Rasoanaivo, Andry},
  booktitle={ICASSP 2020-2020 IEEE International Conference on Acoustics, Speech and Signal Processing (ICASSP)},
  pages={3637--3641},
  year={2020},
  organization={IEEE}
}

@inproceedings{schonfeld2020u,
  title={A u-net based discriminator for generative adversarial networks},
  author={Schonfeld, Edgar and Schiele, Bernt and Khoreva, Anna},
  booktitle={Proceedings of the IEEE/CVF conference on computer vision and pattern recognition},
  pages={8207--8216},
  year={2020}
}

@misc{rui22,
    author = {Rui, Hualan and Beaudoing, Hiroko},
    title = {README Document for NASA GLDAS Version 2 Data Products},
    month = {October},
    year = {2022},
    publisher = {NASA GES DISC}
}

@article{Landerer2012,
author = {Landerer, F. W. and Swenson, S. C.},
title = {Accuracy of scaled GRACE terrestrial water storage estimates},
journal = {Water Resources Research},
volume = {48},
number = {4},
pages = {},
doi = {https://doi.org/10.1029/2011WR011453},
url = {https://agupubs.onlinelibrary.wiley.com/doi/abs/10.1029/2011WR011453},
eprint = {https://agupubs.onlinelibrary.wiley.com/doi/pdf/10.1029/2011WR011453},
year = {2012}
}

@article{Wahr2006,
author = {Wahr, John and Swenson, Sean and Velicogna, Isabella},
title = {Accuracy of GRACE mass estimates},
journal = {Geophysical Research Letters},
volume = {33},
number = {6},
pages = {},
doi = {https://doi.org/10.1029/2005GL025305},
url = {https://agupubs.onlinelibrary.wiley.com/doi/abs/10.1029/2005GL025305},
eprint = {https://agupubs.onlinelibrary.wiley.com/doi/pdf/10.1029/2005GL025305},
year = {2006}
}

@article{Rateb2024,
  title = {Rapid mapping of global flood precursors and impacts using novel five-day GRACE solutions},
  volume = {14},
  ISSN = {2045-2322},
  url = {http://dx.doi.org/10.1038/s41598-024-64491-w},
  DOI = {10.1038/s41598-024-64491-w},
  number = {1},
  journal = {Scientific Reports},
  publisher = {Springer Science and Business Media LLC},
  author = {Rateb,  Ashraf and Save,  Himanshu and Sun,  Alexander Y. and Scanlon,  Bridget R.},
  year = {2024},
  month = jun 
}

@article{Croteau2020,
  title = {Development of a Daily GRACE Mascon Solution for Terrestrial Water Storage},
  volume = {125},
  ISSN = {2169-9356},
  url = {http://dx.doi.org/10.1029/2019JB018468},
  DOI = {10.1029/2019jb018468},
  number = {3},
  journal = {Journal of Geophysical Research: Solid Earth},
  publisher = {American Geophysical Union (AGU)},
  author = {Croteau,  M. J. and Nerem,  R. S. and Loomis,  B. D. and Sabaka,  T. J.},
  year = {2020},
  month = mar 
}

@article{Sun2019,
  title = {Combining Physically Based Modeling and Deep Learning for Fusing GRACE Satellite Data: Can We Learn From Mismatch?},
  volume = {55},
  ISSN = {1944-7973},
  url = {http://dx.doi.org/10.1029/2018WR023333},
  DOI = {10.1029/2018wr023333},
  number = {2},
  journal = {Water Resources Research},
  publisher = {American Geophysical Union (AGU)},
  author = {Sun,  Alexander Y. and Scanlon,  Bridget R. and Zhang,  Zizhan and Walling,  David and Bhanja,  Soumendra N. and Mukherjee,  Abhijit and Zhong,  Zhi},
  year = {2019},
  month = feb,
  pages = {1179–1195}
}

@Article{essd-11-1153-2019,
AUTHOR = {Humphrey, V. and Gudmundsson, L.},
TITLE = {GRACE-REC: a reconstruction of climate-driven water storage changes over the
last century},
JOURNAL = {Earth System Science Data},
VOLUME = {11},
YEAR = {2019},
NUMBER = {3},
PAGES = {1153--1170},
URL = {https://essd.copernicus.org/articles/11/1153/2019/},
DOI = {10.5194/essd-11-1153-2019}
}

@article{lehmann2022well,
  title={How well are we able to close the water budget at the global scale?},
  author={Lehmann, Fanny and Vishwakarma, Bramha Dutt and Bamber, Jonathan},
  journal={Hydrology and Earth System Sciences},
  volume={26},
  number={1},
  pages={35--54},
  year={2022},
  publisher={Copernicus Publications G{\"o}ttingen, Germany}
}

@article{Vishwakarma2018,
  title = {What Is the Spatial Resolution of grace Satellite Products for Hydrology?},
  volume = {10},
  ISSN = {2072-4292},
  url = {http://dx.doi.org/10.3390/rs10060852},
  DOI = {10.3390/rs10060852},
  number = {6},
  journal = {Remote Sensing},
  publisher = {MDPI AG},
  author = {Vishwakarma,  Bramha Dutt and Devaraju,  Balaji and Sneeuw,  Nico},
  year = {2018},
  month = may,
  pages = {852}
}

@misc{https://doi.org/10.48550/arxiv.1809.00219,
  doi = {10.48550/ARXIV.1809.00219},
  url = {https://arxiv.org/abs/1809.00219},
  author = {Wang,  Xintao and Yu,  Ke and Wu,  Shixiang and Gu,  Jinjin and Liu,  Yihao and Dong,  Chao and Loy,  Chen Change and Qiao,  Yu and Tang,  Xiaoou},
  title = {ESRGAN: Enhanced Super-Resolution Generative Adversarial Networks},
  publisher = {arXiv},
  year = {2018},
  copyright = {arXiv.org perpetual,  non-exclusive license}
}

@inproceedings{huang2017densely,
  title={Densely connected convolutional networks},
  author={Huang, Gao and Liu, Zhuang and Van Der Maaten, Laurens and Weinberger, Kilian Q},
  booktitle={Proceedings of the IEEE conference on computer vision and pattern recognition},
  pages={4700--4708},
  year={2017}
}

@misc{https://doi.org/10.48550/arxiv.1905.04899,
  doi = {10.48550/ARXIV.1905.04899},
  url = {https://arxiv.org/abs/1905.04899},
  author = {Yun,  Sangdoo and Han,  Dongyoon and Oh,  Seong Joon and Chun,  Sanghyuk and Choe,  Junsuk and Yoo,  Youngjoon},
  title = {CutMix: Regularization Strategy to Train Strong Classifiers with Localizable Features},
  publisher = {arXiv},
  year = {2019},
  copyright = {arXiv.org perpetual,  non-exclusive license}
}

@article{Wang2023,
  title = {A Review of GAN-Based Super-Resolution Reconstruction for Optical Remote Sensing Images},
  volume = {15},
  ISSN = {2072-4292},
  url = {http://dx.doi.org/10.3390/rs15205062},
  DOI = {10.3390/rs15205062},
  number = {20},
  journal = {Remote Sensing},
  publisher = {MDPI AG},
  author = {Wang,  Xuan and Sun,  Lijun and Chehri,  Abdellah and Song,  Yongchao},
  year = {2023},
  month = oct,
  pages = {5062}
}

@article{Eicker2020,
  title = {Daily GRACE satellite data evaluate short-term hydro-meteorological fluxes from global atmospheric reanalyses},
  volume = {10},
  ISSN = {2045-2322},
  url = {http://dx.doi.org/10.1038/s41598-020-61166-0},
  DOI = {10.1038/s41598-020-61166-0},
  number = {1},
  journal = {Scientific Reports},
  publisher = {Springer Science and Business Media LLC},
  author = {Eicker,  Annette and Jensen,  Laura and W\"{o}hnke,  Viviana and Dobslaw,  Henryk and Kvas,  Andreas and Mayer-G\"{u}rr,  Torsten and Dill,  Robert},
  year = {2020},
  month = mar 
}

@article{Boergens2022,
  title = {Uncertainties of GRACE‐Based Terrestrial Water Storage Anomalies for Arbitrary Averaging Regions},
  volume = {127},
  ISSN = {2169-9356},
  url = {http://dx.doi.org/10.1029/2021JB022081},
  DOI = {10.1029/2021jb022081},
  number = {2},
  journal = {Journal of Geophysical Research: Solid Earth},
  publisher = {American Geophysical Union (AGU)},
  author = {Boergens,  Eva and Kvas,  Andreas and Eicker,  Annette and Dobslaw,  Henryk and Schawohl,  Lennart and Dahle,  Christoph and Murb\"{o}ck,  Michael and Flechtner,  Frank},
  year = {2022},
  month = feb 
}

@article{liwiska2020,
  title = {Preliminary Estimation and Validation of Polar Motion Excitation from Different Types of the GRACE and GRACE Follow-On Missions Data},
  volume = {12},
  ISSN = {2072-4292},
  url = {http://dx.doi.org/10.3390/rs12213490},
  DOI = {10.3390/rs12213490},
  number = {21},
  journal = {Remote Sensing},
  publisher = {MDPI AG},
  author = {Śliwińska,  Justyna and Wińska,  Małgorzata and Nastula,  Jolanta},
  year = {2020},
  month = oct,
  pages = {3490}
}

@article{2019,
  title = {Mass balance of the Greenland Ice Sheet from 1992 to 2018},
  author={{The IMBIE Team}},
  volume = {579},
  ISSN = {1476-4687},
  url = {http://dx.doi.org/10.1038/s41586-019-1855-2},
  DOI = {10.1038/s41586-019-1855-2},
  number = {7798},
  journal = {Nature},
  publisher = {Springer Science and Business Media LLC},
  year = {2019},
  month = dec,
  pages = {233–239}
}

@article{Wouters2008,
  title = {GRACE observes small‐scale mass loss in Greenland},
  volume = {35},
  ISSN = {1944-8007},
  url = {http://dx.doi.org/10.1029/2008GL034816},
  DOI = {10.1029/2008gl034816},
  number = {20},
  journal = {Geophysical Research Letters},
  publisher = {American Geophysical Union (AGU)},
  author = {Wouters,  B. and Chambers,  D. and Schrama,  E. J. O.},
  year = {2008},
  month = oct 
}

@article{Schrama2011,
  title = {Revisiting Greenland ice sheet mass loss observed by GRACE},
  volume = {116},
  ISSN = {0148-0227},
  url = {http://dx.doi.org/10.1029/2009JB006847},
  DOI = {10.1029/2009jb006847},
  number = {B2},
  journal = {Journal of Geophysical Research},
  publisher = {American Geophysical Union (AGU)},
  author = {Schrama,  Ernst J. O. and Wouters,  Bert},
  year = {2011},
  month = feb 
}

@article{scanlon2016global,
  title={{Global evaluation of new GRACE mascon products for hydrologic applications}},
  author={Scanlon, Bridget R and Zhang, Zizhan and Save, Himanshu and Wiese, David N and Landerer, Felix W and Long, Di and Longuevergne, Laurent and Chen, Jianli},
  journal={Water Resources Research},
  volume={52},
  number={12},
  pages={9412--9429},
  year={2016},
  publisher={Wiley Online Library},
  doi={10.1002/2016WR019494}
}


%
%

\end{document}